\documentclass[a4paper,11pt]{article}
\pdfoutput=1
\usepackage{multirow}
\usepackage{jcappub}
\usepackage{amsmath}
\usepackage{bm}
\usepackage{xcolor,color,colortbl}
\usepackage[utf8]{inputenc}
\usepackage{hyperref}
\usepackage{url}
\usepackage{graphicx}
\usepackage{multirow,bigdelim}
\usepackage{amssymb}
\usepackage[normalem]{ulem}
\usepackage[amssymb]{SIunits}
\usepackage{multirow}
\usepackage{bbold}
\usepackage{verbatim}
\usepackage{xspace}

\definecolor{darkgreen}{HTML}{339933}

\graphicspath{{plots/}}

\usepackage{scalerel,tikz}
\usetikzlibrary{svg.path}
\definecolor{orcidlogocol}{HTML}{A6CE39}
\tikzset{orcidlogo/.pic={
 \fill[orcidlogocol] svg{M256,128c0,70.7-57.3,128-128,128C57.3,256,0,198.7,0,128C0,57.3,57.3,0,128,0C198.7,0,256,57.3,256,128z};
 \fill[white] svg{M86.3,186.2H70.9V79.1h15.4v48.4V186.2z}
 svg{M108.9,79.1h41.6c39.6,0,57,28.3,57,53.6c0,27.5-21.5,53.6-56.8,53.6h-41.8V79.1z M124.3,172.4h24.5c34.9,0,42.9-26.5,42.9-39.7c0-21.5-13.7-39.7-43.7-39.7h-23.7V172.4z}
 svg{M88.7,56.8c0,5.5-4.5,10.1-10.1,10.1c-5.6,0-10.1-4.6-10.1-10.1c0-5.6,4.5-10.1,10.1-10.1C84.2,46.7,88.7,51.3,88.7,56.8z};
}}
\newcommand\orcidicon[1]{\href{https://orcid.org/#1}{\mbox{\scalerel*{
\begin{tikzpicture}[yscale=-1,transform shape]
\pic{orcidlogo};
\end{tikzpicture}
}{|}}}}

\title{Selecting samples of galaxies with fewer Fingers-of-God}
\author[a,b,c]{Antón Baleato Lizancos~\orcidicon{0000-0002-0232-6480},}
\author[a,b,c]{Uroš Seljak~\orcidicon{0000-0003-2262-356X},}
\author[a,b]{Minas Karamanis~\orcidicon{0000-0001-9489-4612},}
\author[d,e]{Marco Bonici~\orcidicon{0000-0002-8430-126X},}
\author[a,c]{and Simone Ferraro~\orcidicon{0000-0003-4992-7854}}
\affiliation[a]{Berkeley Center for Cosmological Physics, UC Berkeley, CA 94720, USA}
\affiliation[b]{Department of Physics, University of California, Berkeley, CA 94720, USA}
\affiliation[c]{Lawrence Berkeley National Laboratory, One Cyclotron Road, Berkeley, CA 94720, USA}
\affiliation[d]{Waterloo Centre for Astrophysics, University of Waterloo, Waterloo, ON N2L 3G1, Canada}
\affiliation[e]{Department of Physics and Astronomy, University of Waterloo, Waterloo, ON N2L 3G1, Canada}

\emailAdd{a.baleatolizancos@berkeley.edu}
\emailAdd{useljak@berkeley.edu}
\emailAdd{mkaramanis@berkeley.edu}
\emailAdd{mbonici@uwaterloo.ca}
\emailAdd{sferraro@lbl.gov}

\abstract{The radial positions of galaxies inferred from their measured redshift appear distorted due to their peculiar velocities. We argue that the contribution from stochastic velocities --- which gives rise to `Fingers-of-God' (FoG) anisotropy in the inferred maps --- does not lend itself to perturbative modelling already on scales targeted by current experiments. To get around this limitation, we propose to remove FoG using data-driven indicators of their abundance that are local in nature and thus avoid selection biases. In particular, we show that the scale where the measured power spectrum quadrupole changes sign is tightly anti-correlated with both the satellite fraction and the velocity dispersion, and can thus be used to select galaxy samples with fewer FoG. In addition, we show that excluding galaxies in haloes more massive than a given mass threshold can help to discard many of the most problematic galaxies. Such selection could be achieved in practice using maps of the thermal Sunyaev-Zel’dovich distortion of the cosmic microwave background frequency spectrum. These techniques could potentially improve reconstructions of the large-scale velocity and displacement fields from the redshift-space positions of galaxies. They may also extend the reach of perturbative models for galaxy clustering, though in practice we find only marginal gains when fitting one-loop EFTofLSS models to simulations with mitigated FoG due to the relevance of other effects entering at two-loop order.}

\begin{document}
\maketitle
\flushbottom

\section{Introduction}

By measuring the angular positions and redshifts of galaxies, spectroscopic surveys such as DESI~\cite{DESI} or Euclid~\cite{laureijs2011euclid, Euclid:2024yrr} map the distribution of galaxies on our past lightcone. When converting measured redshifts to radial positions, an error is incurred due to Doppler shifts produced by peculiar velocities being interpreted as cosmological redshifts~\cite{kaiser_clustering_1987}. These redshift-space distortions (RSD) are a powerful cosmological probe in their own right, as they encode information about galaxy velocities sourced by gravitational collapse, with implications for theories of gravity, dark energy and the matter content of the Universe. This information can be retrieved from the anisotropy RSD induce on the observed clustering statistics: at leading order, a projection of the redshift-space power spectrum onto a Legendre basis reveals a valuable signal in the quadrupole and hexadecapole components, not just the isotropic monopole.

Over the last two decades, RSD have been established as an invaluable cosmological probe. Alongside improvements in data quality and volume, developments in theoretical modeling have enabled ever more reliable and efficient predictions against which to contrast the data~\cite{taruya_baryon_2010, seljak_distribution_2011, reid_towards_2011, hand_extending_2017}. One particular family of models, the Effective Field Theory of Large-Scale Structure~\cite{baumann_cosmological_2010, carrasco_effective_2012, vlahLagrangianEffectiveField2015} (EFTofLSS) for biased tracers (see Ref.~\cite{ivanov_effective_2022} for a review), has emerged as the new standard framework thanks to combining the computational efficiency of perturbation theory with the valued property of being a complete and maximally-agnostic basis. The EFTofLSS has recently been used to extract cosmological information from the `full-shape' of galaxy clustering for BOSS~\cite{damico_cosmological_2020, ivanov_cosmological_2020, ivanov_cosmological_2021, ivanov_effective_2022, philcox_boss_2022,chen_new_2022, ivanov_full-shape_2024} and DESI~\cite{collaborationDESI2024VII2024}.

A perturbative treatment of this sort should in principle be able to model the dark matter field out to scales where corrections to the linear-theory calculation of the power spectrum become order unity. In a vanilla $\Lambda$CDM cosmology, this happens at $k_{\mathrm{NL}}\approx0.6\,h$\,Mpc$^{-1}$ for $z=1$ ($k_{\mathrm{NL}}\approx0.3\,h$\,Mpc$^{-1}$  for $z=0$)~\cite{dodelson_12_2021}. However, galaxies are stochastic and biased tracers of the matter distribution~\cite{kaiser_clustering_1987} which significantly complicates the modelling. Current analyses of the two-point clustering of galaxies at $z\sim 1$ using 1-loop EFTofLSS models can typically only model information up to $k_{\rm max}\approx0.2\,h$\,Mpc$^{-1}$~\cite{nishimichi_blinded_2020, maus_analysis_2024}. In part, this must be due to the relevance of higher-loop contributions which are absent from the model, some of which may well be calculable still (see e.g. Ref.~\cite{taule_two-loop_2023}). But perhaps the greatest obstacle to extracting cosmological information from galaxy clustering is the redshift distortions associated with the local dynamics of galaxies in their host halos --- the `Fingers-of-God' (FoG) effect. These velocities are often very large and, crucially, they are random/stochastic: this means they are not correlated with the large-scale velocity field that perturbation theory attempts to model.

The literature is rich in efforts to remove FoG with the aim of better extracting cosmological information. Early works proposed an `FoG-compression' procedure that essentially isotropized the measured redshift-space distribution of galaxies~\cite{tegmark_cosmological_2006, reid_luminous_2009}. The problem with this approach is that the redshift-space position of galaxies is a function of their velocity, 
%and not the dark matter's, 
so the remapping operation introduces a `velocity bias' that is not well understood. Moreover, the non-locality of the selection introduces spurious scale dependence. More recently, alternative measurements have been proposed which discard valuable line-of-sight information~\cite{ivanov_cosmological_2022} or extend the modeling to include some two-loop terms~\cite{damico_taming_2021}. However --- as we will show in this paper --- the contribution from FoG becomes non-perturbative already on rather large-scales, at which point it becomes in principle necessary to model the effect to all orders. The only reason it may at first sight appear as if one-loop models (or perturbative versions of them) have sufficient flexibility to describe the data is that it is only a relatively small fraction of galaxies that are responsible for the most problematic FoG~\cite{schmittfull_modeling_2021}. The corollary to this is that if we could remove those few, most problematic galaxies, the range of validity of the models would be extended with little penalty in terms of signal-to-noise.

In this paper, we propose data-driven ways to select samples of galaxies with fewer FoG while minimizing the impact of selection biases. We identify the zero-crossing of the power spectrum quadrupole as a faithful marker of FoG abundance\footnote{Related ideas were hinted at in footnote 5 of Ref.~\cite{schmittfull_modeling_2021}.} and use simulations to explain this in terms of changes to interpretable halo occupation distribution (HOD) parameters such as the minimum and maximum mass of halos that host galaxies, satellite fractions, or the velocity bias of centrals or satellites relative to their dark matter halo. Our main result in this context is that the scale, $k$, where the quadrupole crosses zero is inversely proportional to the satellite fraction --- cf. figure~\ref{fig:fsat_vs_kcrossing} --- as well as to the velocity dispersion of these galaxies --- cf. figure~\ref{fig:multipole_alpha_dependence}. We then suggest using measurements of the quadrupole to select samples of galaxies with fewer FoG. As an example of immediate applicability, we explore the connection between the power spectrum quadrupole, galaxy color, and FoG; verifying that red galaxies have stronger FoG~\cite{coilDEEP2GalaxyRedshift2008, mohammadVIMOSPublicExtragalactic2018, hangGalaxyMassAssembly2022, mergulhaoEffectiveFieldTheory2023}, and arguing that they should be discarded from a given sample in order to  shift the scale of FoG non-linearity to higher $k$.

We also explore the possibility of removing galaxies living in massive halos, where their thermal velocities are large. This could be done using ancillary thermal Sunyaev-Zel'dovich~\cite{sunyaev_observations_1972} or X-ray data, which has the advatange of having a clean and largely local selection function. We showcase the potential of this approach and emphasize the powerful complementarity between upcoming galaxy redshift surveys and CMB experiments such as the Simons Observatory (SO)~\cite{ade_simons_2019} and CMB-S4~\cite{abazajian_cmb-s4_2016}.

Though we focus on RSD analyses as the main application of our work, we note that being able to identify galaxy samples with fewer FoG also facilitates reconstructions of the large-scale displacement and velocity fields, with far-reaching benefits including for radial baryon acoustic oscillations (BAO) analyses and kSZ velocity reconstruction. Moreover, FoG are typically sourced by satellite galaxies, so samples selected in the ways we propose could help bypass issues to do with miscentering in various applications including tSZ or kSZ stacking.

The structure of this paper is as follows. We begin with an introduction to the theory of redshift-space distortions in \S\ref{sec:intro_to_RSD}. In this section, we also show that the zero-crossing of the power spectrum quadrupole is a useful indicator of the strength of FoG in a given galaxy sample, and argue that this contribution escapes perturbative control already on relatively large scales. In \S\ref{sec:HOD_properties}, we explain what the zero-crossing of the quadrupole is sensitive to by referring to simulations.  The question of whether the EFTofLSS can model the effect of FoG is addressed more quantitatively in \S\ref{sec:assessing_perturbativity}: first, by extracting cosmological constraints from simulations with varying satellite fractions; then --- in order to disentangle FoG from other higher-loop effects --- by analyzing noiseless synthetic data vectors derived from the EFT framework but subjected to anisotropic damping resembling the effect of FoG. This is followed by \S\ref{sec:examples}, where we make contact with applications and argue that discarding the reddest galaxies in a given sample or masking the most massive, tSZ- or X-ray-detected galaxy clusters and halos can be effective ways to mitigate FoG. Finally, we discuss the implications of our work in \S\ref{sec:discussion}.

\section{Redshift-space distortions}\label{sec:intro_to_RSD}
\subsection{Introduction}
The radial position of a galaxy inferred from its redshift (i.e., in `redshift space') differs from its true location due to the Doppler shift associated with its peculiar velocity along the line of sight, which gets interpreted as being due to the background expansion of the Universe. Naturally, this affects the measured clustering statistics. In linear theory, and in the distant observer approximation, the redshift space overdensity of galaxies, $\delta^S$, is related to its real space counterpart, $\delta_g$, via a \emph{Kaiser factor}~\cite{kaiser_clustering_1987}
\begin{equation}
    \delta^S(\bm{k}) = \left( 1 + f \mu^2 /b_1^{E} \right) \delta_g(\bm{k})\,,
\end{equation}
where $f= d \ln D / d \ln a$ is the dimensionless linear growth factor, $b_1^{E}$ is the linear Eulerian galaxy bias and $\mu = \cos \theta $ where $\theta$ is the angle between the wavevector $\bm{k}$ and the line of sight. The measured clustering statistics are therefore a function of the orientation relative to the line of sight, breaking statistical isotropy: the linear theory, redshift-space galaxy power spectrum can be written as
\begin{equation}
    P^S(\bm{k}) = \left( 1 + f \mu^2 /b_1^{E} \right)^2 P_g(\bm{k})\,,
\end{equation}
where we have ommitted any redshift dependence. In addition, 
in linear theory the galaxy power spectrum is related to the dark 
matter power spectrum $P_m(\bm{k})$ via $P_g(\bm{k})=(b_1^{E})^2P_m(\bm{k})$.
We observe that the power spectrum depends both on cosmology via $f$ and $P_m(\bm{k})$ and on the bias parameter $b_1^{E}$. However, because we 
have different angular dependence for different terms one can isolate cosmology information alone in the form of $f^2P_m(\bm{k}) $. To extract 
this information one must measure its angular dependence, or at least two independent angular 
moments, discussed further below.

The Kaiser correction is sourced by the coherent infall of galaxies towards the center of their cluster. This is captured to a large extent by linear theory. However, in addition to the motions induced by these large-scale gravitational potential gradients, there are also contributions associated with the stochastic velocities of galaxies. This contribution, which goes by the name of `Fingers-of-God', is quite pernicious as it is uncorrelated with the large-scale dynamics that perturbation theory models strive to capture. Though a principled model for FoG is not known, a few different phenomenological descriptions have been explored in the literature, achieving remarkable success (e.g.~\cite{sanchezClusteringGalaxiesCompleted2017}). These include multiplying the satellite overdensity by Gaussian or Lorentzian damping profiles (in configuration space)~\cite{peacock_power_1992, percival_testing_2009} of the form
\begin{equation}\label{eqn:damping}
    G(\mu, k\sigma_{v}) =  \exp\left[- (k \sigma_{v} \mu)^2/2\right]\, \quad \text{or} \quad G(\mu, k\sigma_{v}) = 
        \left[1 + (k \sigma_{v} \mu)^2/2\right]^{-1}\,,
\end{equation}
where $\sigma_{v}$ is the velocity dispersion of the galaxy population in question\footnote{Though useful and accurate to a first approximation, these parametrizations are likely to be too simplistic descriptions of reality; see e.g. Ref.~\cite{chen_consistent_2020} for a discussion.}. Typically, it is satellite galaxies living in massive environments that display the largest thermal velocities. Let us therefore split the sample into centrals with subscript $c$ and satellites with subscript $s$, and suppose the fraction of satellites is $f_s$, typically $\mathcal{O}(0.1)$. To simplify the discussion, we will assume that both populations have the same linear bias value, $b_1^{E}$, though generalizing this is trivial. The redshift-space power spectrum can be approximated as
\begin{align}\label{eqn:full_ps}
    P^S(\bm{k}) = \left( 1 + f \mu^2/b_1^{E} \right)^2 \big[ (1-f_s)^2 P_{cc}(k) & + 2f_s(1-f_s)P_{cs}(k) G(\mu, k\sigma_{v}) \nonumber \\
    &+ f_s^2 P_{ss}(k) G^2(\mu, k\sigma_{v})\big]\,,
\end{align}
where for simplicity we have assumed that the centrals move with the center-of-mass of their halo and thus experience no FoG. Some authors further allow for the velocity dispersion to be a function of galaxy environment and mass of the host halo~\cite{okumura_galaxy_2015}. Though quantitatively important, we shall not dwell on these details here, in what is meant to be a heuristic explanation.

In principle, measuring the power spectrum as a function of $\mu$ is requried in order to access all the information at the level of the two-point function. In practice, however, it proves convenient to compress the data into Legendre multipoles,
\begin{equation}\label{eqn:multipoles}
    P_\ell(k) \equiv \frac{2\ell+1}{2}\int_{-1}^{1} d\mu \,\mathcal{L}_\ell(\mu) P^S(\bm{k})\,,
\end{equation}
where $\mathcal{L}_\ell(\mu)$ is the Legendre polynomial of order $\ell$. Most of the signal is contained in the first few multipoles: in linear theory, $\ell=\{0,2,4\}$ contain all the clustering information there is at the two-point level, with the quadrupole in particular carrying most of the cosmology signal. This is because the monopole is dominated by the bias term and on its own contains little cosmology information. It must be supplemented by the quadrupole to 
isolate the cosmological information contained in $f^2P_m(\bm{k}) $. 
A number of insights are readily available from this expression. First, notice that the monopole is essentially an average over $\mu$, so it is dominated by centrals and transverse modes with little FoG effect (or any sort of RSD, for that matter). On the other hand, higher multipoles arise from integrating power spectra against kernels that weight terms in such a way that contributions from central-satellite and satellite-satellite pairs may come to dominate.  We will investigate exactly when this is the case in the next section. Incidentally, the structure of the Legendre polynomials, compounded with the fact that perturbative models are much more accurate for small $\mu$, also means that multipoles with $\ell>0$ are more susceptible to modeling errors --- and FoG --- than all but the most extreme $\mu$-wedges~\cite{chen_consistent_2020}.

\subsection{The zero-crossing of the power spectrum quadrupole}\label{sec:zero_crossing_theory}
Let us consider contributions to the power spectrum in equation~\eqref{eqn:full_ps} of the form $P_{cs}(k) G(\mu, k\sigma_{v})$ and $ P_{ss}(k) G^2(\mu, k\sigma_{v})$, originating from central-satellite and satellite-satellite pairs, respectively. The first pertinent observation is that in the limit that FoG are small and the Kaiser effect dominates, $G(\mu, k\sigma_{v})$ can be linearized, and the redshift-space power spectrum inherits the $k$-dependence of its real-space counterpart, as do its multipoles; i.e the redshift-space power spectrum multipoles look like rescaled versions of the real-space power spectrum, with amplitudes that encode information about the growth rate and the mean matter density (via $f$) as well as galaxy bias. Such linearization is typically a good approximation on large scales (small $k$), for orientations perpendicular to the line of sight (small $\mu$) or for small velocity dispersions. Away from these limits, the effect of non-linear RSD cannot be neglected. 

\begin{figure}
    \centering
    \includegraphics[scale=0.83]{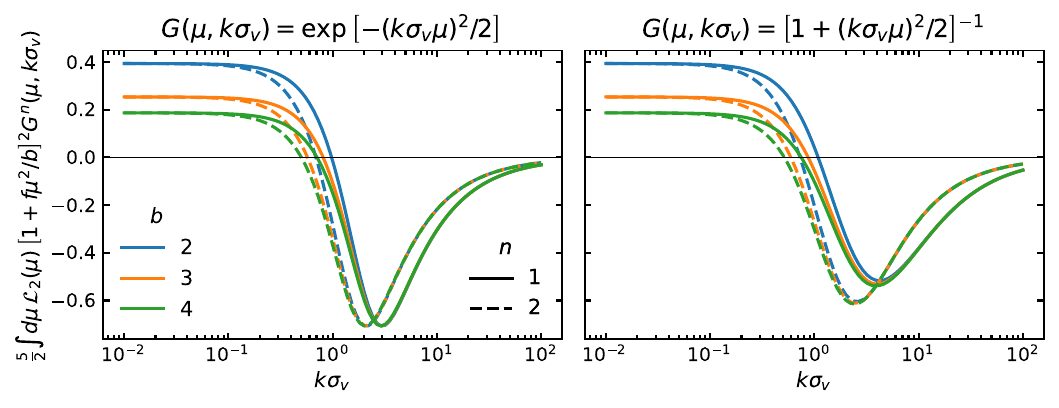}
    \caption{Factor multiplying the real-space power spectrum in the calculation of the redshift-space power spectrum quadrupole of central-satellite ($n=1$) and satellite-satellite ($n=2$) pairs, shown for various values of the linear Eulerian galaxy bias $b$. In the left panel, the FoG are modelled via a Gaussian damping term; on the right, a Lorentzian is used. Both point to the same key takeaway that FoG causes the quadrupole to change sign around $k\sigma\approx 1$.}
    \label{fig:analytic_zerocrossing}
\end{figure}

The nature of the anisotropy induced by RSD is such that most of the additional information it brings in (both linear and non-linear) is contained in the quadrupole. In what follows, we will keep coming back to a particularly useful marker of the onset of non-linearities: the sign change of the power spectrum quadrupole. Following the approximations of equation~\eqref{eqn:full_ps}, the quadropole can be written as
\begin{align}
    P_2(k) = \frac{5}{4} \int_{-1}^{1} d\mu (3\mu^2 - 1)\left[1+f \mu^2/b_1^{E}\right]^2  \big[& (1-f_s)^2 P_{cc}(k) + 2f_s(1-f_s)P_{cs}(k) G(\mu, k\sigma_{v}) \nonumber \\
    &+ f_s^2 P_{ss}(k) G^2(\mu, k\sigma_{v})\big] \,.
\end{align}
Understanding the structure of this integral will be key to assessing in what limit the redshift-space power spectrum can be modeled using perturbation theory. To get there, let us focus on the last two terms, which are the ones that receive contributions from non-linear FoG (since we have assumed that the centrals sit at the center of their host halo and thus move with the dark matter). These take the form
\begin{align}
    P_2(k) \propto P(k) \int_{-1}^{1} d\mu (3\mu^2 - 1)\left[1+f \mu^2/b_1^{E}\right]^2  G^n(\mu, k\sigma_{v}) \,.
\end{align}
where $P(k)$ is now the real-space, central-satellite or satellite-satellite power spectrum for $n=1$ or $n=2$, respectively. Consider the case where $k\sigma_v\gg1$, such that the damping is very effective. The only significant contributions to the integral come from domain regions where $\mu\approx 0$. In these regions $3\mu^2 \ll 1$, so we expect the contribution to the power spectrum quadrupole to be negative --- structures are being elongated along the line of sight, giving rise to `Fingers of God'. By contrast, when the exponential can be linearized, the contribution is positive --- the regime of Kaiser's `pancakes of God'. We see this more quantitatively in figure~\ref{fig:analytic_zerocrossing}, where we have evaluated the integral numerically assuming $f\approx \Omega_{\rm m}^{0.55}$~\cite{peebles_textbook} with the Planck best-fit value of $\Omega_{\rm m}=0.315$~\cite{planck_18_params}.

An important takeaway from figure~\ref{fig:analytic_zerocrossing} is that the sign change happens on scales where $k\sigma_{v}\approx1$; hence, it shifts to larger physical scales the higher the galaxy velocity dispersion. In fact, in this toy model, the $k$ at which the central-central or satellite-central quadrupole crosses zero depends almost exclusively on $\sigma_v$.\footnote{There is also a weak dependence on galaxy bias, as illustrated in Figure~\ref{fig:analytic_zerocrossing}. All other things being equal, higher bias values bring the zero-crossing of the quadrupole to lower $k$.} 

It also follows that the zero-crossing delimits roughly the scales where a perturbative treatment can capture the behavior of these central-satellite and satellite-satellite terms. To see why, consider without loss of generality the Gaussian damping profile. Figure~\ref{fig:analytic_zerocrossing} shows that for bias values typical of satellite galaxies,  $k \sigma_v $ is order unity or larger once the quadrupole goes negative, so in this regime a perturbative expansion of the exponential
\begin{equation}
    e^{-(k \sigma_v \mu )^2} = 1 - (k \sigma_v \mu )^2 + \frac{1}{2!} (k \sigma_v \mu )^4 - \frac{1}{3!} (k \sigma_v \mu )^6 + \dots
\end{equation}
valid for all values of $\mu$ only converges if an infinite number of terms are kept. For reference, a typical satellite velocity dispersion is $\sigma_v\approx6\,h^{-1}\,\mathrm{Mpc}$~\cite{okumura_galaxy_2015} for BOSS CMASS LRGs at $z=0.5$, which corresponds to an FoG non-linearity scale of roughly $k_{\rm{NL};\rm{FoG}}\approx 0.15 \,h\,\mathrm{Mpc}^{-1}$. In fact, non-perturbative effects may manifest themselves already at $k\sigma_v<1$. We will see this later, in \S\ref{sec:eft_mock_dvec}, where we investigate how well cosmological parameters can be recovered from a synthetic data vector featuring the kind of phenomenological FoG damping described here. Without delving into details prematurely, we note that constraints start to become biased already when $k\sigma_v \approx 0.7$ (see figure 9, in particular).

But satellite velocity dispersion does not tell the whole story. Fortunately, the galaxies with strong FoG are typically a limited fraction of the whole sample, and the potentially non-perturbative central-satellite and satellite-satellite contributions to the power spectrum coexist with a central-central term that is easier to model. One of the main goals of this paper is to show that all other things being equal, samples for which the power spectrum quadrupole crosses zero at higher $k$ have either a lower proportion of satellites or lower velocity dispersions. Consequently, these samples will have more limited non-perturbative contributions from FoG. In the following section, we investigate halo-occupation statistics in $N$-body simulations to more rigorously identify what properties of the sample might be changing when the zero-crossing scale shifts.

\section{What sample properties affect the zero-crossing of the quadrupole?}\label{sec:HOD_properties}
In the previous section, we identified the zero-crossing of the power spectrum quadrupole as an indicator of FoG strength in a given galaxy sample. We now set out to explore what sample properties may be underlying this behavior. We do this by populating the \texttt{AbacusSummit} $N$-body simulations with galaxies following various halo occupation distribution (HOD) prescriptions and measuring their clustering statistics.

\subsection{Simulations}\label{sec:abacus}
We use $N$-body simulations from the \texttt{AbacusSummit} suite~\cite{maksimova_span_2021}. In particular, we use the \texttt{base} volume realizations, of which 25 different realizations at the Planck 2018 cosmology\footnote{This is a $\Lambda$CDM cosmology with parameters $\Omega_c h^2 = 0.1200$, $\Omega_b h^2 = 0.02237$, $\sigma_8 = 0.811355$, $n_s = 0.9649$, $h = 0.6736$, $w_{0}=-1$, and $w_a= 0$.} are provided. Each simulated box contains $6912^3$ particles within a $(2 \,h^{-1}\,\rm{Gpc})^3$ volume, resulting in a particle mass of $2.1 \times 10^9 \,h^{-1} \,M_{\odot}$. Altogether, the 25 realizations have a total volume of $200\,(h^{-1}\,\rm{Gpc})^3$, about 35 times that of the CMASS and LOWZ samples from BOSS DR12, the combination of which covers a volume of 5.7 $(h^{-1}\,\rm{Gpc})^3$~\cite{dawson_baryon_2013} which in turn is about half of the volume covered by the DESI LRG samples~\cite{collaboration_desi_2024}. We choose to work with the snapshot at $z=0.5$ so that we can ignore redshift evolution or selection effects and focus on the impact of non-linear RSD.

Following \cite{yuan_span_2022}, we `paint' galaxies onto these dark-matter-only simulations with properties similar to the BOSS CMASS sample using a halo occupation distribution (HOD). Tools to do this are provided as part of the \texttt{AbacusHOD} module of \texttt{AbacusUtils}\footnote{https://abacusutils.readthedocs.io/en/latest/index.html}. In the next section we go into the details of the various HODs we explore. The reader interested in technical details about our power spectrum measurements or covariance matrix calculations may consult appendices~\ref{appendix:pk_measurements} and~\ref{appendix:mock_covariances}, respectively.

\subsection{HOD model}
We use the baseline model implemented in \texttt{AbacusHOD}~\cite{yuan_span_2022}, which extends the five-parameter model of Ref.~\cite{zheng_galaxy_2007} to include central and satellite velocity bias as well as incompleteness. In this model, the probability of a halo of mass $M$ hosting a central galaxy follows a binomial distribution with mean
\begin{equation}
	\bar{n}_{\rm cen}(M) = \frac{1}{2}\mathrm{erfc}\bigg[ \frac{\log_{10}(M_{\rm cut}/M)}{\sqrt{2}\sigma} \bigg]\,,
\end{equation}
whereas for satellites the distribution is Poissonian with mean
\begin{equation}\label{eqn:sat_hod}
	\bar{n}_{\rm sat}(M) = \bigg[ \frac{M - \kappa M_{\rm cut}}{M_1}\bigg]^{\alpha} \bar{n}_{\rm cen}(M)  \,.
\end{equation}
In this parametrization, $M_{\rm cut}$ determines the minimum mass a halo must have before it can host a central galaxy; $\sigma$ characterizes the slope of the transition between the regimes where halos have zero or one galaxy; $M_1$ controls the typical mass of halos that host a single satellite galaxy; $\alpha$ is the power-law index controlling the mean number of satellites as a function of halo mass; and finally, $\kappa M_{\rm cut}$ sets the suppression of satellite abundance in the low-mass end.

In this flavor of \texttt{AbacusHOD}, the central galaxy is placed at the center of mass of the largest subhalo and assigned its velocity vector. On the other hand, satellites  are assigned to particles of the halo, with equal probability. The framework allows for a `velocity bias' following Ref.~\cite{guoVelocityBiasSmall2015} that introduces deviations between the velocities of centrals and satellites and the dark matter particles they are assigned to. For centrals, this is modelled as an additional Gaussian scatter of the line-of-sight velocity,
\begin{equation}\label{eqn:velocity_bias_central}
    v_{\rm c} = v_{\rm halo} + \alpha_c\, \delta v(\sigma_{\rm LOS})\,,
\end{equation}
where $v_{\rm halo}$ is the line-of-sight velocity of the halo; $\delta v(\sigma_{\rm LOS})$ is the Gaussian scatter; and $\alpha_c$ is the centrals' velocity bias parameter: $\alpha_c=0$ if the central moves with the center of mass of the halo. For satellites, on the other hand, the model is
\begin{equation}\label{eqn:velocity_bias}
    v_{\rm s}  = v_{\rm halo} + \alpha_s\, (v_{\rm{s}, 0} - v_{\rm halo})\,,
\end{equation}
where $v_{\rm{s}, 0}$ is the line-of-sight velocity of the dark matter particle --- i.e., the satellite velocity when $\alpha_s=1$ in which case there is no velocity bias.

Finally, the HOD model also includes a completeness fraction $f_{\mathrm{ic}}$ to account for sample selection effects~\cite{leauthaudStripe82Massive2016, guoIncompleteConditionalStellar2018}. Unless it is otherwise stated, we will use the best-fit value of $f_{\mathrm{ic}}=0.58$ determined by Ref.~\cite{yuan_span_2022}; we have checked that varying this parameter --- for example to $f_{\rm ic}=1$ --- does not affect the conclusions of this section in any significant way. Later on, when trying to constrain cosmological parameters from the mocks in \S\ref{sec:assessing_perturbativity}, we will set $f_{\rm ic}=1$ in order to simplify the modeling.

\subsection{Results}\label{sec:hod_results}
In \S\ref{sec:zero_crossing_theory}, we used a toy model to argue that the zero-crossing of the quadruple is a marker of FoG strength. Let us now verify this by establishing a connection between this observable and interpretable HOD parameters which we know are closely related to stochastic galaxy velocities.

A priori, we expect parameters such as $M_1$, $\alpha$ and $\kappa$ to play potentially important roles in this regard. We therefore focus on how varying them affects the power spectrum multipoles. Table~\ref{tab:fsat_values_diffHODs} shows the parameter ranges we consider: $M_1$ and $\alpha$ are set to the mean and $\pm 1\,\sigma$ of the Gaussian prior used in Ref.~\cite{yuan_span_2022} when fitting to BOSS CMASS; for $\kappa$, on the other hand, we consider the best-fit value of $\kappa=0.2$ from the same reference and two more extreme values of 1.0 and 5.0 (their prior is centered at $\kappa=0.5$ with width $\pm 2$). 

\begin{figure}
    \centering
    \includegraphics[scale=0.72]{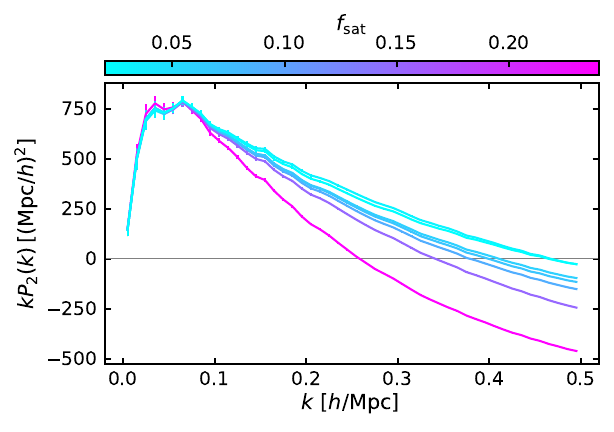}
    \includegraphics[scale=0.72]{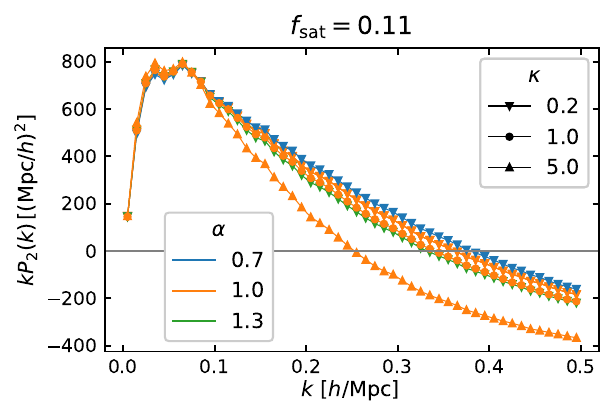}
    \caption{Power spectrum quadrupoles measured from simulations built around HOD models with values of $M_1$, $\alpha$ and $\kappa$ given in table~\ref{tab:fsat_values_diffHODs}. Different models lead to different values of $f_{\rm sat}$. In the left panel, where galaxies are weighted equally, changes in the zero-crossing are largely driven by such changes to $f_{\rm sat}$, which depends strongly on $M_1$ (from bottom to top, the curves correspond to the rows of table~\ref{tab:fsat_values_diffHODs}). In the right panel, satellites and centrals are weighted in such a way that $f_{\rm sat}=0.11$ in all cases. All other things being equal (including the satellite fraction), higher values of $\alpha$ mean that satellites populate relatively higher-mass halos, which induce larger thermal velocities and thus have zero-crossings at lower $k$. Raising $\kappa$ at a fixed $f_{\rm sat}$ has a qualitatively similar effect. On the other hand, $M_1$ has no additional effect once we fix $f_{\rm sat}$, and the two curves where we vary it lie under the $\{\alpha=1.0, \kappa=0.2\}$ curve. Measurements are averaged over all 25 simulated boxes, and error bars are drawn from the diagonal of the disconnected covariance matrix appropriate for each HOD model, scaled to the appropriate effective volume. \label{fig:quadrupole_M1_alpha_dependence}}
\end{figure}

As it turns out, most of the effect of these HOD parameters can be condensed to the satellite fraction they give rise to --- naturally, different $f_{\rm sat}$ translate to different amounts of FoG, as the satellite velocities tend to be dominated by the stochastic component. The $f_{\rm sat}$ associated with each of the parameter combinations is given in table~\ref{tab:fsat_values_diffHODs}, and the impact on the measured power spectrum quadrupoles is shown in the left panel of figure~\ref{fig:quadrupole_M1_alpha_dependence}. This figure suggests that changes to the zero-crossing are driven by changes to $f_{\rm sat}$, which in turn is strongly correlated with $M_1$ since this controls the minimum mass at which halos start getting satellites; higher values of $M_1$ tend to place satellites in higher-mass halos, which are strongly suppressed as we approach the high-mass tail of the halo mass function.

Beyond modifications of the satellite fraction, there may also be other physical reasons why these models display different amounts of RSD non-linearities. In order to isolate these effects, we weight particles in such a way that $f_{\rm sat}$ remains fixed at a value of our choosing. More concretely, we apply the same weight to all satellites, and a different but also constant weight to all centrals. Given a satellite fraction, $f^{\rm old}_{\mathrm{sat}}$, intrinsic to the mocks, and a new one, $f^{\rm new}_{\mathrm{sat}}$, we want to enforce, the weights of satellites and centrals are related as
\begin{equation}\label{eqn:weights}
    w_{\rm s} = w_{\rm c} \left(\frac{f^{\rm new}_{\mathrm{sat}}}{1-f^{\rm new}_{\mathrm{sat}}}\right)\left(\frac{f^{\rm old}_{\mathrm{sat}}}{1-f^{\rm old}_{\mathrm{sat}}}\right)^{-1} \,.
\end{equation}
Since the value of $w_{\rm c}$ is arbitrary, we are free to set it to $w_{\rm c}=1$. An alternative approach to fixing $f_{\mathrm{sat}}$ could have been to subsample satellites or centrals depending on whether the model over-produces one or the other. However, our approach has the advantage of not discarding any galaxies, thus offering better statistics as well as preserving all contributions from central-satellite and satellite-satellite pairs.

The right panel of figure~\ref{fig:quadrupole_M1_alpha_dependence} shows the power spectrum quadrupoles measured from mocks populated with the different HOD models of table~\ref{tab:fsat_values_diffHODs} after reweighting the galaxies to enforce $f_{\rm sat}=0.11$ (the best-fit value for BOSS CMASS in the base model of Ref.~\cite{yuan_span_2022}). After normalizing all models to the same $f_{\rm sat}$, the impact of $M_1$ becomes negligible. We do however see a trend where higher values of either $\alpha$ or $\kappa$ shift the zero-crossing to lower $k$. This is because HODs with higher values of these parameters tend to place satellites (of which we now have a `fixed' number) in higher-mass halos; in turn, these have deeper gravitational potentials that source stronger thermal velocities and thus more FoG. Nevertheless, the dependence on these parameters is weaker than the dependence on $f_{\rm sat}$, with most zero-crossings clustering around similar values of $k$ except for the very extreme (and unphysical) model with $\kappa=5$. 
% It is also worth noting that the monopole changes a lot less through these modifications to the HOD at fixed $f_{\rm sat}$ --- changes which, at any rate, should be much more amenable to perturbative modelling. 
%
\begin{table}[h!]
\centering
\begin{tabular}{|c|c|c|c|}
\hline
$\log_{10} (M_{1}/h^{-1}\,M_{\odot})$ & $\alpha$ & $\kappa$ & $f_{\rm sat}$ \\
\hline
\hline
13.80 & 1.00 & 0.2 & 0.24 \\ 
\hline
14.30 & 0.70 & 0.2 & 0.15 \\ \hline
14.30 & 1.00 & 0.2 & 0.09 \\ 
\hline
14.30 & 1.00 & 1.0 & 0.07 \\ 
\hline
14.30 & 1.30 & 0.2 & 0.06 \\ \hline
14.80 & 1.00 & 0.2 & 0.03 \\ \hline
14.80 & 1.00 & 5.0 & 0.02 \\ \hline
\end{tabular}
\caption{Average satellite fractions resulting from varying HOD parameters $M_{1}$, $\alpha$ and $\kappa$ amongst the values shown. The corresponding power spectrum quadrupoles measured from these simulations are shown in the left panel of figure~\ref{fig:quadrupole_M1_alpha_dependence}. The other HOD parameters are set to the best-fit values obtained by Ref.~\cite{yuan_span_2022} when fitting to BOSS CMASS, i.e. $\log_{10} (M_{\rm cut}/h^{-1}\,M_{\odot})=12.86$, $\log_{10} \sigma =-2.8$, $\alpha_{\rm c}=0.22$, $\alpha_{\rm s}=0.98$ and $f_{\rm ic}=0.58$.}
\label{tab:fsat_values_diffHODs}
\end{table}

\begin{figure}
    \centering
    \includegraphics[scale=0.72]{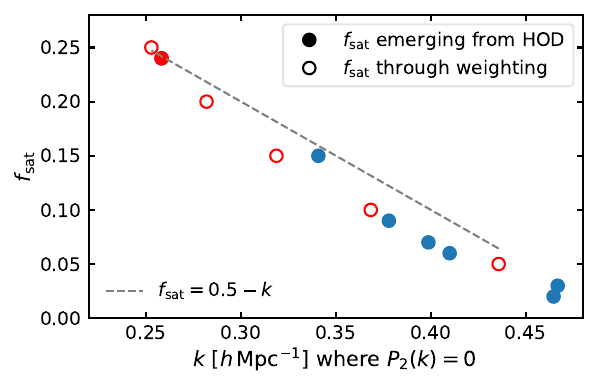}
    \caption{Relationship between the satellite fraction, $f_{\rm sat}$, and the $k$ at which the power spectrum quadrupole crosses zero. Filled circles denote measurements from the different HODs of table~\ref{tab:fsat_values_diffHODs}, while empty, red ones are the result of differentially weighting satellites and centrals in the specific HOD model marked with a red, filled point in order to give rise to some desired $f_{\rm sat}$. A very similar relation holds in both cases: we mark with a dashed line a rough approximation to it (not a fit).\label{fig:fsat_vs_kcrossing}}
\end{figure}

If we plot the $k$ at which the quadrupole crosses zero against the satellite fraction for all of these HOD models --- figure~\ref{fig:fsat_vs_kcrossing} --- we see a clear anti-correlation between the two variables with relatively little scatter. The relation appears in fact to be the same whether these two variables are measured directly from the catalogs produced by each of the HOD models in table~\ref{tab:fsat_values_diffHODs}, or as a result of differentially weighting satellites and centrals in any one particular model to adjust the effective $f_{\rm sat}$. For the CMASS-like LRGs we are working with, the relation can be described (roughly) as $f_{\rm sat} \approx 0.5 - k_{\rm crossing}$.

\begin{figure}
    \centering
    \includegraphics[scale=0.74]{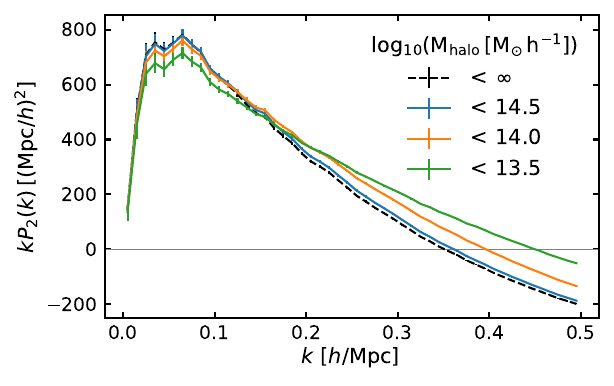}
    \caption{Excising the most massive halos shifts the zero-crossing of the quadrupole --- and thus the scale of FoG non-linearity --- to higher $k$. The reason for this can be understood further from figure~\ref{fig:masking_tsz}, which shows that the galaxies with the highest velocity residual tend to live in the most massive halos.}\label{fig:masking_massive_halos}
\end{figure}

Our discussion around figure~\ref{fig:quadrupole_M1_alpha_dependence} also suggests that the most massive halos will host the galaxies with the highest peculiar velocities. Moreover, we also know that the number of satellites in a given halo grows rapidly with its mass --- cf. equation~\eqref{eqn:sat_hod}. We thus expect that removing the most massive halos should help alleviate FoG. This is indeed what we see in figure~\ref{fig:masking_massive_halos}, where we plot the power spectrum quadrupoles after removing galaxies in halos above three different halo masses: $\{10^{13.5}, 10^{14.0}, 10^{14.5}\}\,M_{\odot}\,h^{-1}$. It is worth noting that varying these HOD parameters naturally brings about changes to the galaxy bias; see, for example, the changes at low $k$ in figure~\ref{fig:masking_massive_halos}. However, these effects are much more likely to be amenable to a perturbative description via the EFTofLSS than FoG are.

\begin{figure}
    \centering
    \includegraphics[scale=0.74]{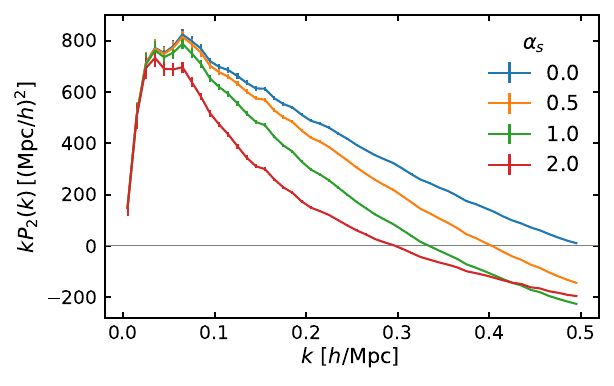}
    \includegraphics[scale=0.74]{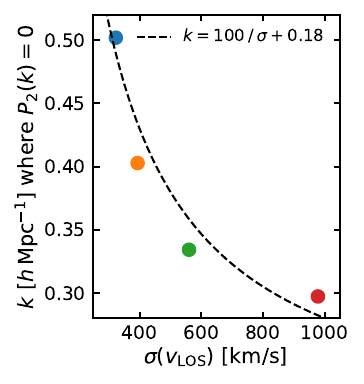}
    \caption{Power spectrum quadrupole (left panel) of a CMASS-like HOD model after rescaling the satellite velocities to various values of the `velocity bias' of equation~\eqref{eqn:velocity_bias}. On the right panel, we show that the zero-crossing is inversely proportional to the velocity dispersion of the satellites arising from each of these models, and overlay an approximate relation between the two variables.\label{fig:multipole_alpha_dependence}}
\end{figure}

It is also instructive to investigate the impact of galaxy peculiar velocities directly. In order to mitigate or accentuate the Fingers-of-God effect, we adjust the speed of satellites along the line of sight  using the satellite velocity bias model in equation~\eqref{eqn:velocity_bias}, leaving the central velocity bias fixed at the fiducial value of $\alpha_c=0.22$, and fixing also $\log_{10} (M_{1}/h^{-1}\,M_{\odot}) = 14.10$ and $\log_{10} (M_{\rm cut}/h^{-1}\,M_{\odot}) = 12.86$.\footnote{See also figure~A4 of Ref.~\cite{avila_completed_2020} for the impact of velocity bias on the power spectrum multipoles of eBOSS emission-line galaxies.} The resulting satellite fraction is $f_{\rm sat}=0.11$. The left panel of figure~\ref{fig:multipole_alpha_dependence} shows the power spectrum quadrupole, averaged over the 25 realizations, for various values of $\alpha_{s}$. As expected, higher values of $\alpha_{s}$ shift the zero-crossing to lower $k$, as they correspond to increasing the radial velocity of the galaxies%\footnote{Notice that the $k$ of zero-crossing reaches a minimum when $\alpha=2$ and then appears to increase slightly for $\alpha=3$. This can be understood from figure~\ref{fig:analytic_zerocrossing}: the most negative contribution to the quadrupoles happens `shortly' after the zero crossing, and then the factor asymptotes back to zero.  Note also that we might be seeing a change in shape as the quadrupole approaches the shape of the monopole due to the damping term going to zero (and the monopole being dominated by central-central pairs).}
. In the panel on the right, we compare the $k$ of zero-crossing to the RMS line-of-sight velocity dispersion of the satellites, highlighting the inverse proportionality between the two. This is in qualitative agreement with our analytic explorations around figure~\ref{fig:analytic_zerocrossing} where we learned that the typical structure of FoG damping is such that $\sigma(v_{\rm LOS})\propto 1/k_{\rm crossing}$.

\section[Assessing the perturbativity of Fingers-of-God (FoG)]{Assessing the perturbativity of Fingers-of-God}\label{sec:assessing_perturbativity}

Before giving practical examples of how to select samples of galaxies with fewer Fingers-of-God, let us motivate why we set out to do so. We will focus, in particular, on whether FoG can be modeled using the EFTofLSS in `full-shape' analyses of the redshift-space power spectrum of galaxies. To this end, we begin in \S\ref{sec:breakdown_of_Pk} by using the HOD infrastructure of the previous section to understand the relative contribution of satellites and centrals to the observed clustering statistics as a function of scale. Then, in \S\ref{sec:eft_tests}, we will show that one-loop EFTofLSS models can fail to describe contributions from FoG even after certain two-loop terms designed to capture them are included.

\subsection[Breakdown of contributions to $P_{\ell}(k)$]{Breakdown of contributions to $P_{\ell}(k)$}\label{sec:breakdown_of_Pk}
According to our discussion around figure~\ref{fig:analytic_zerocrossing}, the contribution from central-satellite and satellite-satellite pairs potentially becomes non-perturbative when the quadrupole of their power spectrum goes negative. In order to determine exactly what scale this corresponds to in either case, and also to understand the relative importance of different contributions to the measured multipoles, we will use the \texttt{Abacus} mocks again, populated with galaxies using a fixed HOD model --- concretely, the `baseline' model of Ref.~\cite{yuan_span_2022} with the best-fit value from analyzing BOSS CMASS data; the only changes we introduce is to set the completeness to $f_{\mathrm{ic}}=1$ and to set the velocity bias to zero to avoid additional modeling difficulties\footnote{All in all, the HOD parameters are the following: $\log_{10} (M_{1}/h^{-1}\,M_{\odot})=14.10$, $\kappa=0.2$, $\alpha=1.12$, $\log_{10} (M_{\rm cut}/h^{-1}\,M_{\odot})=12.86$, $\log_{10} \sigma =-2.8$, $\alpha_{\rm c}=0$, $\alpha_{\rm s}=1$ and $f_{\rm ic}=1$.}. This time, however, we will use the weights in equation~\eqref{eqn:weights} to disentangle contributions from central-central, central-satellite and satellite-satellite pairs.

\begin{figure}
    \centering
    \includegraphics[scale=0.74]{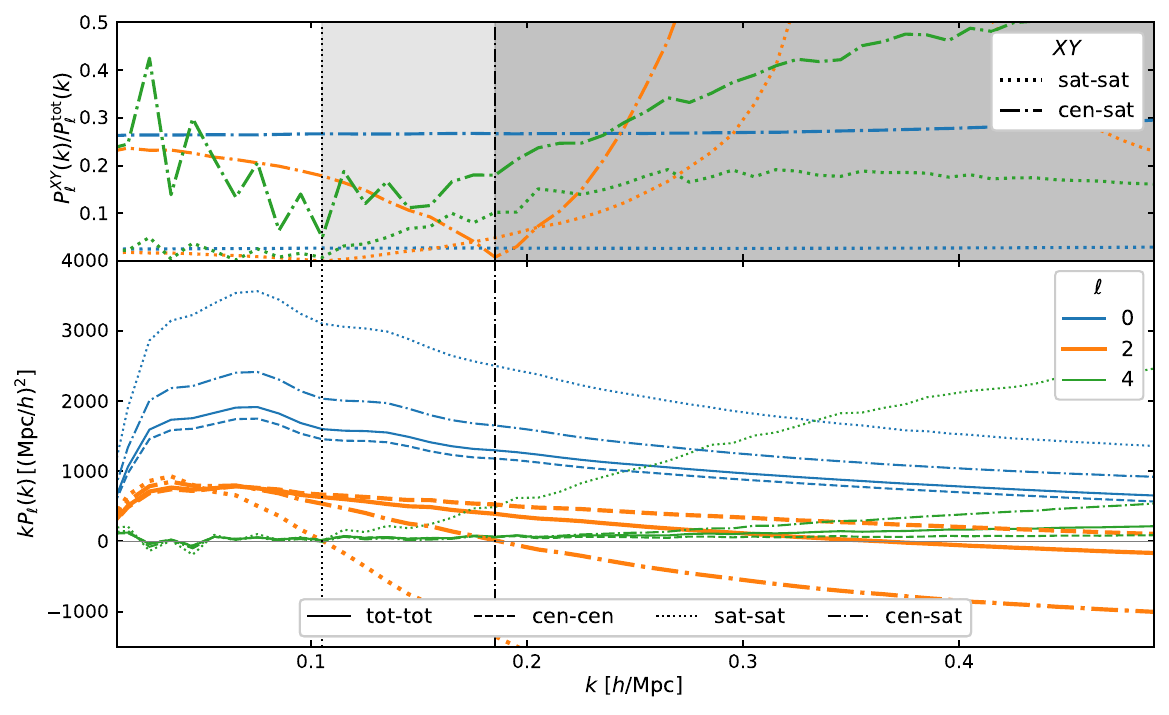}
    \caption{\emph{Lower panel:} Power spectrum multipoles of different pairings of satellites and centrals for \texttt{Abacus} mocks populated with an HOD to resemble BOSS CMASS galaxies. To recover contributions from different pairings to the power spectrum of the total sample (solid), spectra must be multiplied by $(1-f_{\rm sat})$ or $f_{\rm sat}$ for every factor of centrals or satellites, respectively. The important point is that  satellite-satellite and central-satellite contributions are expected to become non-perturbative when their $P_{2}(k)<0$; this scale is marked with vertical dotted and dot-dashed lines for the two respective cases that cut across (and apply to) both panels. \emph{Upper panel:} The fractional contribution to the measured power spectrum multipoles from central-satellite (dot-dashed) and satellite-satellite (dotted) pairs. For simplicity, we assume that the centrals have no thermal velocities so they are always amenable to a perturbative description.}\label{fig:breakdown_Pk_abacus}
\end{figure}

The lower panel of figure~\ref{fig:breakdown_Pk_abacus} shows the power spectrum of all the possible pairings of centrals and satellites (to get their contribution to the power spectrum of the whole sample, one needs to scale every factor of the centrals by $1-f_{\rm sat}$ and every factor of the satellites by $f_{\rm sat}$). The key point to emphasize here is that the quadrupole becomes negative around $k\approx0.11\,h\,\rm{Mpc}^{-1}$ for satellite-satellite pairs and $k\approx0.19\,h\,\rm{Mpc}^{-1}$ for central-satellite pairs\footnote{On the other hand, for central-central pairs, the quadrupole is always positive across the range shown, as in this HOD model the centrals are stationary relative to the center of mass of their halo.}. Our hypothesis is that on these scales and smaller the contributions these make to the measured power spectrum are presumably non-perturbative.

A pertinent question is: how large are those contributions? The upper panel of figure~\ref{fig:breakdown_Pk_abacus} attempts to answer this by showing the amplitude of terms that are potentially non-perturbative --- i.e., the central-satellite power above $k>0.19\,h\,\rm{Mpc}^{-1}$  and the satellite-satellite power above $k>0.11\,h\,\rm{Mpc}^{-1}$ --- relative to the total power spectrum multipoles. The first thing to note is that, although the satellite-satellite contribution is supposed to escape perturbativity early on, the fact that $f_{\rm sat}\sim 0.1$ limits inaccuracies to at most 5\% per mode for $\ell=0$ and $\ell=2$, and 10\% for $\ell=4$ in the regime where the central-satellite contributions are still perturbative --- that is, below $k\lesssim0.19\,h\,\rm{Mpc}^{-1}$. Beyond that, inaccuracies become much larger. The monopole suffers an approximately $k$-independent 25\% contribution per mode from non-perturbative terms at $k\gtrsim0.19\,h\,\rm{Mpc}^{-1}$, while non-perturbative corrections to the quadrupole and hexadecapole grow much quicker and become $\mathcal{O}(50\%)$ around $k\approx 0.25\,h\,\rm{Mpc}^{-1}$. Concerning as these estimates are, we note that they may actually be overly optimistic: in \S\ref{sec:eft_mock_dvec}, below, we will study cosmological parameter recovery in the presence of a phenomenologial FoG damping factor, and find that parameters are biased already at $k\sigma_v\approx0.7$ ($k$ slightly smaller than the zero-crossing scale) when $f_{\rm sat}=1$.

Next, we will build on this particular example to assess the extent to which the EFTofLSS can handle FoG and extract unbiased cosmological constraints when they are present. In order to do that, let us first introduce this modeling framework.

\subsection[EFTofLSS constraints on cosmology in the presence of FoG]{EFTofLSS constraints on cosmology in the presence of FoG}\label{sec:eft_tests}

\subsubsection[Introduction to the EFTofLSS model]{Introduction to the EFTofLSS model}\label{sec:eft_model}
The EFTofLSS program is an extension of decades of developments in cosmological perturbation theory. The latter comes in two flavors: Eulerian PT (EPT), which focuses on the evolution of density and velocity fields; and Lagrangian PT (LPT) which keeps track of the displacement of particles labeled by their initial positions. 
We will follow the latter formalism in this explanation, though both approaches can be shown to be equivalent at every order~\cite{chen_consistent_2020, chen_redshift-space_2021, maus_comparison_2024}.

The Lagrangian approach to EFT treats the dark matter field as a collisionless fluid that only acquires an effective viscosity due to the backreaction of small scale physics on the large-scale dynamics. The initial (Lagrangian) particle positions, $\bm{q}$, are related to the final (Eulerian) ones by $\bm{x}(\bm{q}) = \bm{q} + \Psi(\bm{q})$, where the non-linear displacement field, $\Psi$, is obtained by perturbatively solving the equation
\begin{equation}\label{eqn:psi_eq}
    \partial^2_{\tau} \bm{\Psi} + \mathcal{H} \partial_{\tau} \bm{\Psi} = - \nabla_{\bm{x}} \Phi (\bm{q} + \Psi)\,,
\end{equation}
where $\mathcal{\tau}$ is the conformal time. On the right-hand side, we have the gravitational potential, $\Phi$, which is sourced by the matter overdensity, $\delta_{\rm m}$, via Poisson's equation. In LPT, this differential equation is solved order-by-order in powers of the initial density field, $\delta^{(0)}$. Once the displacements are determined, particles are advected from their initial to their final positions following
\begin{equation}
    1 + \delta_m (\bm{x}) = \int d^3 \bm{q} \delta_{\rm D} \left(\bm{x} - \bm{q} - \Psi(\bm{q})\right)\,.
\end{equation}

To account for the fact that galaxies are biased tracers of the dark matter and baryon distribution~\cite{kaiser_spatial_1984}, the galaxy overdensity can be written as a functional of the matter density field (see e.g.~\cite{chen_redshift-space_2021})
\begin{equation}\label{eqn:deltag_lpt}
    1 + \delta_g (\bm{k}) = \int d^3 \bm{q} e^{i\bm{k}\cdot (\bm{q}+\Psi(\bm{q}))} F[\delta^{(0)}(\bm{q})]\,.
\end{equation}
The only restriction on the bias functional, $F[\delta^{(0}(\bm{q})]$, is that it respect the symmetries of the physical laws involved in galaxy formation: statistical isotropy and homogeneity, as well as the equivalence principle; as such, the approach is maximally conservative. To third order in the linear overdensity --- the order relevant for a typical `one-loop' calculation of the power spectrum --- the real-space galaxy field can include terms of the form
\begin{equation}
    F[\delta^{(0)}(\bm{q})] = 1 + b_1 \delta_0 + \frac{1}{2}b_2 \left(\delta_0(\bm{q})^2 - \langle \delta_0^2\rangle  \right) + b_s \left(s_0^2(\bm{q}) - \langle s_0^2\rangle\right) + b_3 \mathcal{O}_3(\bm{q})\,,
\end{equation}
where $b_1, b_2, b_s$ and $b_3$ are bias parameters to be determined from either the data or simulations and $s_0=\left(\partial_i \partial_j / \partial^2 -  \delta_{i j }/3\right) \delta_0$ is the initial shear tensor. Note that we have written the cubic bias contribution schematically as $\mathcal{O}_3$; unless stated otherwise, in this work we will set $b_3=0$ as this parameter is constrained to be small for typical samples~\cite{abidi_cubic_2018}.

The expression above holds in real space. However, the position of galaxies is redshift space is set not only by the real space displacements, but also --- and crucially for the purposes of this work --- by the distortions induced by the proper motions of galaxies along the line of sight. The redshift-space displacements can be obtained as a boost $\Psi_s = \Psi + \dot{\Psi}$. In an Einstein-de Sitter Universe, this has perturbative solutions $\Psi_s^{(n)} = \Psi^{(n)} + \left(\Psi^{(n)} \cdot \bm{v}\right) f \hat{n}$. This turns out to also be a very good approximation in $\Lambda$CDM, and it can be evaluated efficiently as a coordinate transformation via matrix multiplication (e.g.,~\cite{chen_new_2022}).

From the redshift-space displacements, one can finally obtain a model for the galaxy power spectrum. Let us skip ahead momentarily and quote the final prediction before returning and addressing the different terms in turn. To one-loop order, the redshift-space galaxy power spectrum can be written as\footnote{Note that we are using the reparametrization of the counterterms in Ref.~\cite{maus_analysis_2024}, as this better lends itself to interpreting parameter values as fractional corrections to linear theory.}
\begin{align}\label{eqn:velocileptors_model}
    P_s(\bm k) = &\, P^{\rm LPT}_s \nonumber \\
    & + k^2 (b_1^{E}+f\mu^2) (b_1^{E} \alpha_0 + f \alpha_2 \mu^2 + f^2 \alpha_4 \mu^4) P_{\rm{lin}}(\bm{k}) \nonumber \\
    & + \mathrm{SN}_0 + \mathrm{SN}_2 k^2 \mu^2 + \mathrm{SN}_4 k^4 \mu^4\,.
\end{align}
The first of these terms, $P^{\rm LPT}_s$, is the calculation of the `loops' using perturbation theory. In LPT, this can be obtained directly from equation~\eqref{eqn:deltag_lpt} as
\begin{equation}\label{eqn:P_LPT}
    P^{\rm LPT}_s (\bm{k}) = \int d^3\bm{q} \bigg\langle  e^{i\bm{k}\cdot (\bm{q}+\Delta_s)} F(\bm{q}_1) F(\bm{q}_2)\bigg\rangle_{\bm{q}=\bm{q}_1 - \bm{q}_2}\,,
\end{equation}
where $\Delta_s = \Psi_s(\bm{q}_1) - \Psi_s(\bm{q}_2) $ are the pairwise displacements. This equation is to be solved using the cumulant theorem --- see Ref.~\cite{chen_redshift-space_2021} for details. This formulation conveys clearly the notion of `IR resummation' of long-wavelength modes, which in LPT can proceed simply by keeping linear displacements below some $k_{\rm{IR}}$ exponentiated while Taylor-expanding smaller-scale fluctuations. In this work, however, we will follow an alternative approach that resembles more closely the fiducial choices of the recent DESI full-shape analysis~\cite{collaborationDESI2024VII2024}. This entails calculating the loop contributions using the `Resummed Eulerian Perturbation Theory' (REPT) infrastructure in the \texttt{velocileptors} code\footnote{\url{github.com/sfschen/velocileptors}}~\cite{chen_consistent_2020, chen_redshift-space_2021}, which collects terms in a manner analogous to LPT but carries out the `IR resummation' for both displacements and velocities using the wiggle/no-wiggle split. This approach requires a remapping of Lagrangian bias parameters into their Eulerian version, for which we follow e.g.~\cite{chen_consistent_2020, maus_comparison_2024}. Any values we quote for bias or other nuisance parameters will be in the Lagrangian context. The last ingredient in this calculation is the linear power spectrum of the cold dark matter plus baryon fluid, which we obtain from \texttt{CAMB}~\cite{Lewis:1999bs} or \texttt{CLASS}~\cite{lesgourgues_cosmic_2011} codes, finding indistinguishable results between the two.

The key to the EFT approach is to split the behavior of large- and small-scale modes, integrating away results coming from small-scale physics below the chosen cutoff  while taking into account the back-reaction of this procedure on larger scales~\cite{baumann_cosmological_2010, carrasco_effective_2012}. It can be shown that the small scales contribute a number of counterterms and stochastic contributions: these are the additional terms shown in the second and third lines above. The counterterms are those multiplying the linear matter power spectrum, $P_{\rm{lin}}(\bm{k})$ in the second line. Some authors have advocated for the inclusion of an additional counterterm scaling as $\alpha_{(k\mu)^4}(k \mu)^4 P_{\rm lin}$ --- technically a two-loop correction --- in order to better capture FoG~\cite{damico_taming_2021}; we will study the efficacy of this approach where relevant. 

On the other hand, the last line in equation~\eqref{eqn:velocileptors_model} above contains `stochastic' terms meant to capture small-scale physics uncorrelated with the large-scale dynamics: anything arising from correlations of stochastic velocity and density modes, such as Fingers-of-God (which are also degenerate with the counterterms described above) or shot noise. Using physical insights --- see e.g. Ref.~\cite{maus_comparison_2024} --- priors can be placed on this these parameters.

The size of the counterterms relative to the linear theory prediction can be a useful indicator of the breakdown of perturbativity~\cite{damicoLimitsPrimordialNonGaussianities2022, bragancaPeekingNextDecade2023,zhangHODinformedPriorEFTbased2024}: when they become comparable, the perturbative hierarchy is not to be trusted. In this regime, we can also conclude that the cosmological information has been saturated, since this is extracted primarily from comparing measurements to calculations of the linear theory and the loop integrals, while the counterterms and stochastic contributions mostly parametrize contributions that are uncorrelated with the large-scale density fluctuations~\cite{chen_consistent_2020}.

\subsubsection{An analysis of Abacus mocks with varying satellite fractions}\label{sec:abacus_fits}
Having explained the EFT model in the previous section, let us now put it to the test by extracting cosmological constraints from the CMASS-like simulations described in \S\ref{sec:breakdown_of_Pk}. By differentially weighting satellites and centrals, we are able to isolate their contributions to the power spectrum, which we have argued are impacted to different extents by FoG. We will now fit the EFT model to these measurements in order to gauge whether it has sufficient flexibility to accurately account for the FoG or, alternatively, whether the cosmological constraints we extract are biased.

According to Bayes' theorem, the posterior probability of the parameters given the data is proportional to the product of prior probability and data likelihood. We use the $k_{\rm max}$-dependent priors of table~\ref{tab:priors}, which are inspired by the fiducial choices of DESI~\cite{collaborationDESI2024VII2024}, particularly with regards to the counterterm and stochastic parameters\footnote{We depart slightly from the DESI choices by sampling directly the bias parameters rather than rescaled versions of the form $b_1\sigma_8, b_2 \sigma_8^2$, etc. While the latter parametrization can help reduce prior volume effects, this is not expected to be an issue for us given the large volumes we use; indeed we have checked in certain, specific cases that our results are consistent across both approaches. Our priors on cosmological parameters are also different --- we use a restricted range in order to reduce the cost of training our emulator --- but see no evidence of this affecting our results.}. The likelihood, on the other hand, involves our measured data vector, a covariance matrix and a model prediction. For computational efficiency, we build a neural-network emulator for the \texttt{velocileptors} model predictions using the \texttt{Effort}~\cite{boniciEffortFastDifferentiable2025} framework (see appendix~\ref{appendix:emulator}). This model prediction includes Alcock-Paczynsky distortions (AP; see appendix~\ref{sec:AP}) and a $k$-rebinning scheme to account for the discreteness of the simulation box (see appendix~\ref{appendix:discreteness}).

Finally, we sample the posterior using \texttt{pocoMC}\footnote{We find satisfactory convergence when using 4096 effective particles, half as many active ones, dynamic particle allocation and t-preconditioned Crank-Nicolson MCMC. We also find that using a normalizing flow preconditioner significantly helps characterize the posteriors. \texttt{pocoMC} is available at: \url{https://pocomc.readthedocs.io/en/latest/}}~\cite{karamanis2022pocomc,karamanis_accelerating_2022}, fixing the parameters that redshift-space clustering does not constrain well to the true values underlying the \texttt{Abacus} simulations, namely $\omega_b=0.02237, n_s=0.9649, N_\nu=2.033, \tau=0.0568, M_\nu=0.06$. We then sample the cosmological parameters $\{H_0,\omega_c, \ln(10^{10} A_{\rm s})\}$ and different sets of nuisance parameters depending on whether or not we fit the hexadecapole. When we do,  we vary the bias parameters $\{b_1, b_2, b_s, b_3\}$ as well as the counterterm and stochastic contributions $\{\alpha_0, \alpha_2, \alpha_4, \rm{SN}_0, \rm{SN}_2, \rm{SN}_4\}$. When we do not, we fix $\alpha_4=\mathrm{SN}_4=0$; additionally, we set $b_3=0$ in order to match the baseline choices of DESI, and also because this parameter is small and not well constrained by the data. In either case, we also consider varying the $\alpha_{(k\mu)^4}$ counterterm to assess whether it helps mitigate FoG.

\begin{figure}
    \centering
\includegraphics[scale=0.77]{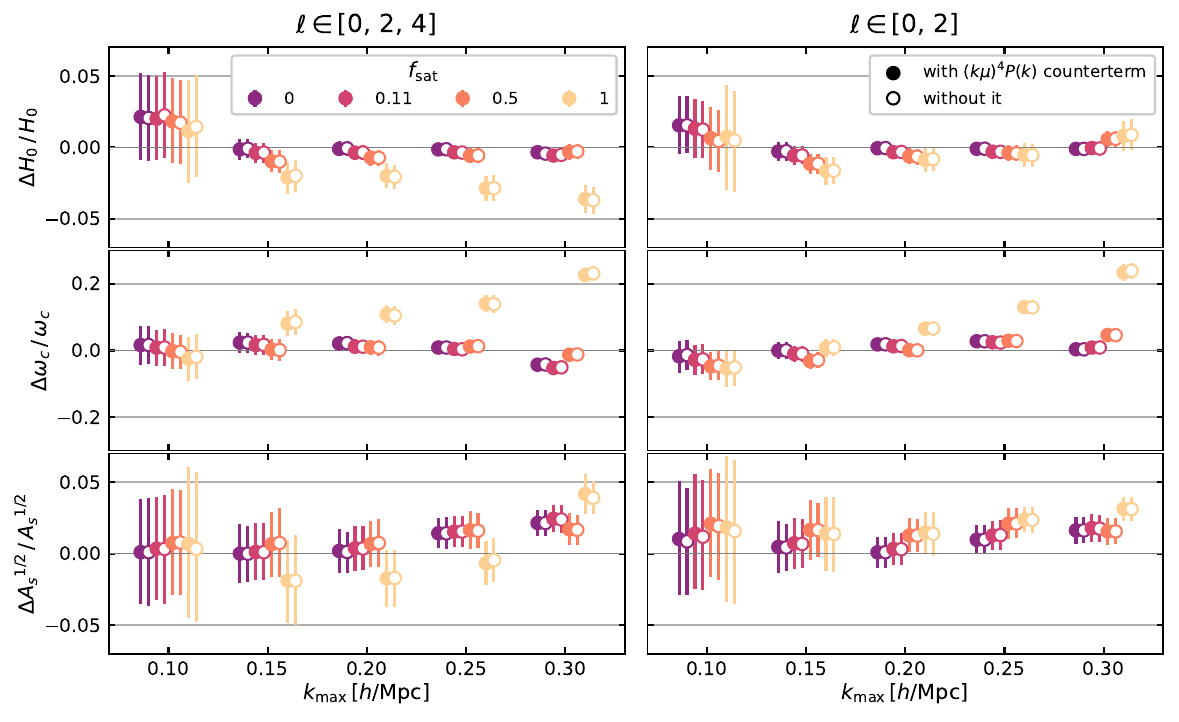}
    \caption{Marginalized posterior mean and 68\% credible intervals on cosmological parameters as a function of $k_{\rm max}$ obtained from analyzing \texttt{Abacus} simulations populated with BOSS CMASS-like galaxies (as described in \S\ref{sec:breakdown_of_Pk}). (The points at a given given $k_{\rm max}$ are shifted for visualization purposes.) The right panel shows fits to the monopole and quadrupole while the left panel shows results including the hexadecapole as well; in the latter case, three more parameters are varied: $\alpha_4$, $\mathrm{SN}_4$ and $b_3$. Filled (empty) points correspond to varying (fixing to zero) the $\alpha_{(k\mu)^4}$ counterterm. Different colors correspond to different values of the satellite fraction, which we isolate through differentially weighting satellites and centrals in the mock catalogs. Although the modeling breaks down even without any satellites ($f_{\rm sat}=0$), it is clear that at any given $k_{\rm max}$, a higher satellite fraction leads to more biased cosmological inference. Adding the $\alpha_{(k\mu)^4}$ counterterm does not mitigate these effects significantly.}\label{fig:cosmo_constraints_abacus}
\end{figure}

Figure~\ref{fig:cosmo_constraints_abacus} shows our marginalized constraints on cosmological parameters as a function of the $k_{\rm max}$ employed in the fits. For a given data vector and $k_{\rm max}$, there is a clear trend where as  $f_{\rm sat}$ increases, constraints get more biased relative to the known truth underlying the simulations. Including the hexadecapole seems to lead to larger biases: while the increase is marginal for moderate values of $f_{\rm sat}$, the breakdown is rather dramatic when $f_{\rm sat}=1$, for which constraints on $\omega_c$ are biased at the 10\% level already at $k_{\rm max} = 0.15\,h\,\rm{Mpc}^{-1}$. Including the additional $\alpha_{(k\mu)^4}$ counterterm does not significantly mitigate the biases on the inferred cosmology.

Since the satellite fraction correlates strongly with FoG strength, these results illustrate the difficulty of accounting for FoG in perturbative models such as the EFTofLSS. However, the reader will have noted that even in the case where we have no satellites (and no peculiar velocity of the centrals relative to the halo)
, cosmological constraints become biased by $k_{\rm max} = 0.25\,h\,\rm{Mpc}^{-1}$. This is in qualitative agreement with earlier work in Ref.~\cite{ivanov_cosmological_2021} which showed, using $N$-body simulations populated with emission-line galaxies following an HOD, that lowering the satellite fraction leads to only slightly more accurate constraints on cosmological parameters with no significant gains in terms of the $k$-reach of the one-loop theory. This could well be because higher-loop contributions independent of FoG are becoming relevant. Unlike FoG, these corrections may well be perturbative; for example, Ref.~\cite{taule_two-loop_nodate} showed recently that two-loop corrections to the redshift-space matter power spectrum at $z=0.5$ are significant already at $k=0.15\,h\,\rm{Mpc}^{-1}$. In the next section, we set out to disentangle the impact of FoG from these other, higher-loop effects.

\subsubsection{An analysis of a manifestly perturbative data vector acted on by FoG}\label{sec:eft_mock_dvec}

In this section we investigate to what extent the effect of FoG can be captured by the EFTofLSS parametrization. We have argued above that tests on $N$-body simulations do not lend themselves to disentangling the effect of FoG from other (potentially perturbative) effects coming in at two-loop order and beyond. To get around this issue and isolate the effect of FoG alone, we now test the redshift-space EFT model's ability to get unbiased cosmological constraints given a `composite', synthetic data vector obtained by applying a phenomenological FoG damping factor to a noiseless redshift-space power spectrum drawn from the EFT model itself: that is, we construct a synthetic data vector following
\begin{equation}\label{eqn:composite_pkmu}
    P^S(\bm{k}) = (1-f_s)^2 P^{S}_{cc}(\bm{k}) + 2f_s(1-f_s)P^{S}_{cs}(\bm{k}) G(\mu, k\sigma_{v}) + f_s^2 P^{S}_{ss}(\bm{k}) G^2(\mu, k\sigma_{v})\,,
\end{equation}
with $P^{S}_{cc}(\bm{k})$, $P^{S}_{cs}(\bm{k})$ and $P^{S}_{ss}(\bm{k})$ drawn from the EFT model itself. In this formulation, the redshift-space space power spectrum is a combination of the redshift-space auto- and cross- power spectra of satellites and centrals, appropriately weighted by the satellite fraction and scaled by phenomenological damping factors. We will use a Lorentzian parametrization of the damping factor with a conservative $\sigma_v=7\,h^{-1}\,\rm{Mpc}$, and a fiducial satellite fraction of 11\% derived from the HOD described in \S\ref{sec:breakdown_of_Pk}, which is inspired by BOSS CMASS.

We determine the parameters going into the `perturbative' part of the mock power spectra --- i.e., the redshift-space EFT model --- by fitting the `composite' model (with fixed damping factors) to the central-central, central-satellite and satellite-satellite power spectrum multipoles of the CMASS-like \texttt{AbacusHOD} mock described in \S\ref{sec:breakdown_of_Pk} and shown in figure~\ref{fig:breakdown_Pk_abacus}. During the fitting process, we keep the cosmological parameters fixed to the ground truth underlying the simulation and use scales up to $k_{\rm{max}}=0.25\,h^{-1}\,\rm{Mpc}$. Every draw of our model is multiplied by an `observation matrix' that accounts for discreteness effects in the \texttt{Abacus} boxes and a rebinning in $k$; see appendix~\ref{appendix:discreteness}.

Procedurally, we do this fitting by defining a likelihood function under the assumption that the data residuals are Gaussian-distributed given some model parameters and using a covariance matrix obtained from simulations following appendix~\ref{appendix:abacus_appendix}. For these kinds of `full-shape' analyses, the likelihood is a high-dimensional and non-Gaussian function of the model parameters given the data, thus requiring robust optimization methods. We use simulated annealing, an algorithm for finding global optima through stochastic sampling. Heuristically speaking, this method avoids getting stuck in local maxima by introducing a non-zero probability of accepting samples in regions of lower likelihood. This probability is controlled by a global parameter, $\beta$, which behaves much like the (inverse) thermodynamic temperature of a system. Early on, the effective temperature is high so the probability of accepting these `worse' steps is relatively large, but as $\beta$ increases and the system cools (it is crucial that it does so smoothly) the probability drops steadily, such that towards the final iterations the process asymptotes to a `greedy algorithm' that only climbs uphill towards the local maximum\footnote{More practically, we rescaled the log-likelihood by the $\beta$ parameter and progressively raised the latter according to $\beta\propto T_i^3$ where $T_i$ consist of 200 evenly-spaced steps in $T_i$. At each iteration, we sample the log-likelihood function through MCMC until the maximum of the likelihood corresponding to our sampled points stops increasing. It is possible to speed up convergence by regularizing the objective function so as to introduce an effective Gaussian prior on the parameters with widths and means of our choosing. We then move on the next iteration at a slightly lower temperature, and sample again using the previous best-fit points as the initial guess. To improve efficiency, we discard the half of the samples that have the lower likelihood, and respawn an equal number of particles at the locations of the remaining ones, only with a small perturbation to the parameter values. The choice of initial guess for the parameters can also speed up convergence significantly: we find that $\mathcal{O}(10^{-3})$ perturbations about the input EFT parameters works sufficiently well for the case in hand}.

We fit the monopole, quadrupole and hexadecapole moments of the power spectrum using a restricted-freedom scenario where we vary only $\{b_1, b_2, b_s, \alpha_0, \alpha_2, \rm{SN}_0, \rm{SN}_2\}$ and set the remaining EFT parameters to zero. Typically, $\alpha_4$ and $\rm{SN}_4$ are also varied when fitting to the hexadecapole, but we do not do so here in order to ensure that the resulting synthetic data vector is amenable to a perturbative description in the absence of FoG even when $\alpha_4=\rm{SN}_4=0$, as is typically assumed when fitting only to the monopole and quadrupole (e.g.~\cite{collaborationDESI2024VII2024}).

As part of the optimization, we use a regularization scheme enforcing that the best-fit parameters lie within one standard deviation of the central value of the priors in table~\ref{tab:priors}. The best-fit counterterms are thus restricted to always be a small correction of the linear theory prediction. Restricting the parameter freedom this way obviously degrades the fit quality, but we emphasize that our goal is only to qualitatively reproduce the relative contributions from central-central, central-satellite and satellite-satellite pairs; there is no reason to expect the fit to be highly accurate (even after folding in the non-perturbative damping factor), given that, as we have argued, the \texttt{Abacus} mocks receive significant contributions on these scales from two-loop terms besides FoG.
\begin{table}[h]
\centering
    \begin{tabular}{||c||}
    \hline
     Cosmology\\
    \hline
    \hline
    $H_0$\\
    $\mathcal{U}[60, 74]$ \\
    \hline
    $\omega_b$\\
    $\mathcal{U}[0.08, 0.16]$ \\
    \hline
    $\ln (10^{10}\,A_s)$\\
    $\mathcal{U}[3.4, 2.6]$ \\
    \hline
    \hline
    \end{tabular}
    \quad
    \begin{tabular}{||c||}
        \hline
            Bias\\
            \hline
            \hline
            $b_1$ \\
            $\mathcal{U}[-1, 3]$ \\
            \hline
            $b_2$ \\
            $\mathcal{N}[0, 8]$ \\
            \hline
            $b_s$ \\
            $\mathcal{N}[0, 8]$\\
            \hline
            $b_3$ \\
            $\mathcal{N}[0, 8]$\\
            \hline
            \hline
    \end{tabular}
    \quad
    \begin{tabular}{||c||}
        \hline
         Counterterms \& Stochastic\\
        \hline
        \hline
        $\alpha_0$ \\
        $\mathcal{N}[0, 12.5 \times (0.2/k_{\rm{max}})^2]$ \\
        \hline
        $\alpha_2$\\
        $\mathcal{N}[0, 12.5 \times (0.2/k_{\rm{max}})^2]$\\
        \hline
        $\alpha_4$\\
        $\mathcal{N}[0, 12.5 \times (0.2/k_{\rm{max}})^2]$\\
        \hline
        $\alpha_{(k\mu)^4}$\\
        $\mathcal{N}[0, 12.5 \times (0.2/k_{\rm{max}})^2]$\\
        \hline
        $\rm{SN}_0$\\
        $\mathcal{N}[0, 2/\bar{n}]$ \\
        \hline
        $\rm{SN}_2$\\
        $\mathcal{N}[0, 5f_{\rm{sat}} \sigma_{v}^2 /\bar{n}]$ \\
        \hline
        $\rm{SN}_4$\\
        $\mathcal{N}[0, 5f_{\rm{sat}} \sigma_{v}^4 /\bar{n}]$ \\
        \hline
        \hline
    \end{tabular}
    \caption{Priors used in our Bayesian analyses. Here, $\mathcal{U}(a,b)$ is a uniform distribution from $a$ to $b$, while $\mathcal{N}(a,b)$ is a Gaussian with standard deviation $b$ centered at $a$. $\bar{n}$ is the number density of galaxies, set to $\bar{n}=5\times10^{-4}$ when not determined from mocks; $\sigma_v$ is the satellite velocity dispersion, for which we use $\sigma_v=7h^{-1}\rm{Mpc}$; and $f_{\rm sat}$ is the satellite fraction, set as appropriate for each scenario.
    }
    \label{tab:priors}
\end{table}

The optimization procedure is carried out independently for central-central and satellite-satellite pairs, yielding the best-fit parameters of table~\ref{tab:redshift_space_model}. The best-fit parameters from these auto-spectra then fix all the nuisance parameters of the cross-spectrum except for the stochastic contributions; this prediction has been worked out in the `multitracer' literature~\cite{mergulhao_effective_2022, mergulhaoEffectiveFieldTheory2023, ebina_cosmology_2024}. We therefore hold the bias and counterterm parameters fixed to the multi-tracer predictions when maximizing the likelihood of the central-satellite pairs, and vary only the stochastic parameters. For consistency, we fix $\rm{SN}_4=0$ for the cross-spectrum, just like we did for the two auto-spectra. Optimization tells us that the remaining two stochastic terms in the cross-spectrum have best-fit values $\rm{SN}_0 = 1107\pm6$ and $\rm{SN}_2 = -45600\pm800$. All in all, this fitting procedure ensures that the central and satellite contributions to each multipole of our synthetic data vector are in rough agreement with figure~\ref{fig:breakdown_Pk_abacus} and the \texttt{Abacus} mocks.

\begin{figure}
    \centering
\includegraphics[scale=0.74]{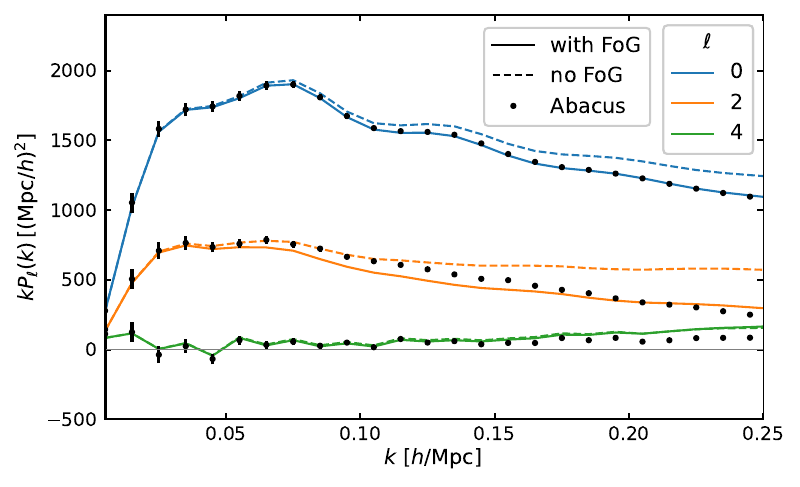}
    \caption{Redshift-space power spectrum multipoles derived from an EFT model with parameters given in table~\ref{tab:redshift_space_model} before (dashed) and after (solid) applying a damping factor that mimics the effect of FoG. To facilitate comparison with the \texttt{Abacus} mock that we are trying to qualitatively match (black points with error bars) the spectra are multiplied by an `observation matrix' that accounts for discreteness effects (hence the jagged structure at low $k$). We are interested in seeing whether these mock data can be fit by the EFT framework after the FoG damping has been applied.}
    \label{fig:mock_pk}
\end{figure}

We now use the central value of these best-fit parameters to draw new central-central, central-satellite and satellite-satellite spectra from the EFT model and scale them by FoG damping factors as required; finally we linearly combine the terms weighting each one of them by the appropriate factors of $f_{\rm sat}$. The resulting power spectrum multipoles for the full sample --- extracted from the redshift-space power spectrum of equation~\eqref{eqn:composite_pkmu} using~\eqref{eqn:multipoles} --- are shown in figure~\ref{fig:mock_pk}. We also compare them to measurements on the \texttt{Abacus} mocks, seeing that our fitting ensures good qualitative agreement. Our synthetic data vector thus contains a mixture of central and satellite contributions that is representative of the truth (insofar as the simulations are realistic), but crucially has no two-loop contributions besides those introduced by FoG.

\begin{table}[h]
    \centering
    \resizebox{\textwidth}{!}{%
    \begin{tabular}{||c||c|c|c|c|c|c|c||}
        \hline
           & $b_1$ & $b_2$ & $b_s$ & $\alpha_0$  & $\alpha_2$  & $\rm{SN}_0$ & $\rm{SN}_2$ \\
         \hline
         \hline
         Centrals & $0.818 \pm 0.006$ & $-2.4 \pm0.1$ & $0.2 \pm 0.1$ & $8 \pm 1$ & $-8 \pm 2$ & $-1500 \pm 200$ & $-5000 \pm 1000$ \\ \hline
         Satellites & $1.71 \pm 0.02$ & $0.6 \pm0.4$ & $-1.0 \pm 0.5$ & $2 \pm 3$ & $8 \pm 8$ & $4000 \pm 1000$ & $-182000 \pm 6000$ \\
         \hline
    \end{tabular}
    }
    \caption{Best-fit parameters from fitting the central-central and satellite-satellite power spectra in figure~\ref{fig:breakdown_Pk_abacus} (obtained from an $N$-body+HOD mock) with a composite model featuring a perturbative (EFT) component and \emph{ad hoc} FoG damping with satellite velocity dispersion  $\sigma_v=7h^{-1}\rm{Mpc}$. These parameters fix the central-satellite cross-spectrum up to stochastic contributions, which we determine to be $\rm{SN}_0 = 1562\pm4$ and $\rm{SN}_2 = -27200\pm 200$. The uncertainties we quote are derived from the information matrix and are merely orientative. When combining central-central, satellite-satellite and central-satellite contributions into a total data vector, we will use a satellite fraction of $f_{\rm sat}=0.11$ as measured from the reference HOD mock.}
    \label{tab:redshift_space_model}
\end{table}

\subsubsection*{Can the EFTofLSS parametrize a phenomenological FoG damping term?}\label{sec:constr_sats_only_toy}
The anisotropic suppression of clustering induced by FoG can be described phenomenologically by the Gaussian or Lorentzian damping factors of equation~\eqref{eqn:damping}~\cite{okumura_galaxy_2015}. These analytic forms let us argue that, for central-satellite or satellite-satellite pairs, $k\sigma_v>1$ when $P_{2}(k)<0$, suggesting that the clustering of such pairs cannot be described perturbatively beyond the zero-crossing of the quadrupole.

One way to investigate this more quantitatively is to ask whether the EFTofLSS model can fit the phenomenological damping factor when the cosmological parameters are fixed to the known truth. We explore this route in appendix~\ref{appendix:fitting_damping}, finding that given seven EFT parameters, one can fit the damping factor quite accurately even beyond the zero-crossing scale. However, the best-fit values of the parameters often lie in regions of parameter space that are disfavored by typical priors (and in the case of counterterms, deemed unphysical on the grounds of perturbativity). Moreover, since the EFT and cosmological parameters are correlated, there is no guarantee that inferences of the latter will remain unbiased once they are freed up.

The more relevant question is therefore whether cosmological constraints are biased when FoG are present. This subsection addresses this point in the limit that our synthetic data vector is made up exclusively of satellites (we will consider the `full sample' in the next subsection). To do this, we will once again sample the posterior probability on the parameters given the synthetic data, using the priors in table~\ref{tab:priors}, with only slight modifications to the likelihood described in \S\ref{sec:abacus_fits}. We once again use the \texttt{Effort} emulator for the \texttt{velocileptors} model, including AP distortions. Although the data vector is known with no uncertainty, it is still important to assign the correct relative weight between different $k$'s when fitting. We thus use an analytic disconnected covariance described in appendix~\ref{appendix:analytic_gauss_covmat}, assuming there is no shot noise in our measurements. The high accuracy of the \texttt{Effort} emulator and the fact that our data vector is noiseless allow us to rescale the covariance to an effective volume of $1000\,\rm{Gpc}^3$ while keeping emulator error limited to at most $0.5\,\sigma$ at each $k$. Working with such a large effective volume lets us avoid projection effects and dramatically reduces the uncertainty in our determination of the inference biases associated with FoG. Also, with such a large volume, the prior becomes relatively unimportant~\cite{hadzhiyskaCosmology6Parameters2023}, and changes to it do not significantly impact our results.

\begin{figure}
    \centering
\includegraphics[scale=0.77]{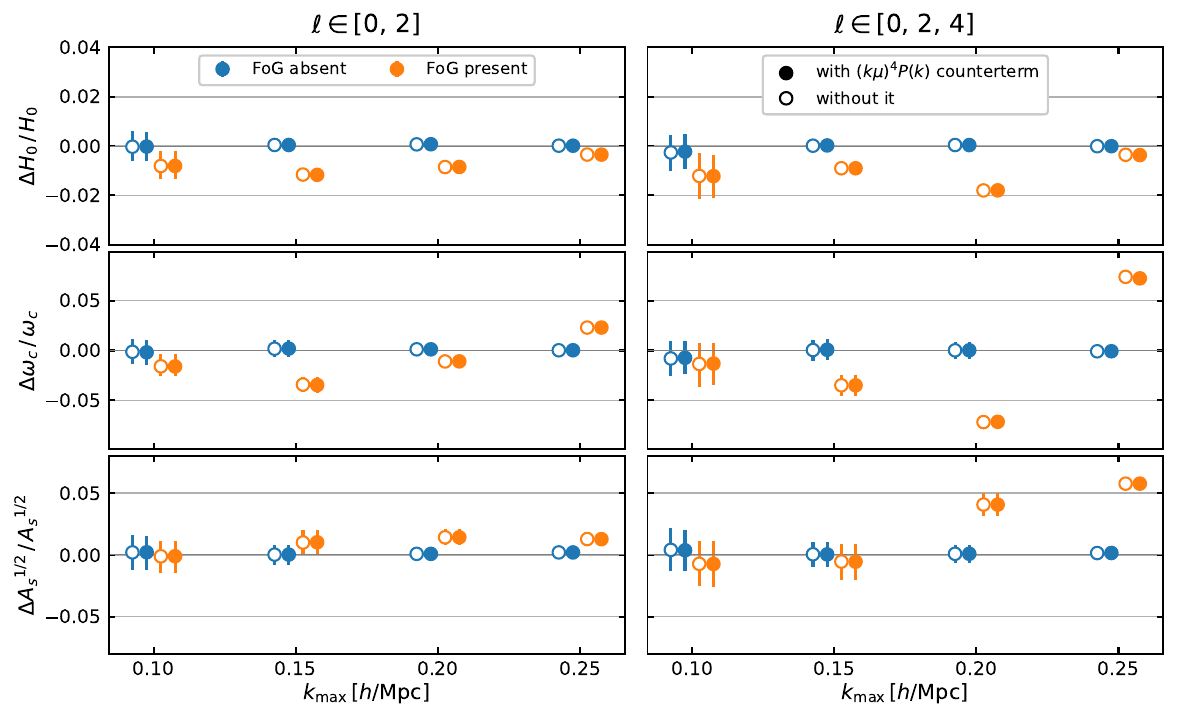}
    \caption{Marginalized posterior mean and 68\% credible intervals on cosmological parameters as a function of $k_{\rm max}$, obtained from analyzing a noiseless, composite EFT+damping synthetic data vector for which $\sigma_v=7h^{-1}\rm{Mpc}$ and $f_{\rm sat}=1$. (A small horizontal shift is applied to points at the same $k_{\rm max}$ for visualization purposes.) When no FoG damping is applied (blue) the data is manifestly perturbative and constraints are unbiased; but when a phenomenological damping resembling FoG is applied (orange) the inferred cosmology becomes biased. Adding an extra counterterm which is formally of two-loop order (empty markers) does not amelliorate the situation relative to the case where it is not included (filled markers). The left panel shows fits to the monopole and quadrupole, while the right one includes the hexadecapole as well (in which case we vary also $\alpha_4, \rm{SN}_4$ and $b_3$). Constraints are plotted as a fractional deviation from the true parameter value underlying the simulations.}
    \label{fig:constraints_sats_only}
\end{figure}

Figure~\ref{fig:constraints_sats_only} shows the marginalized constraints on cosmological parameters $H_0, \omega_c$ and $A_s$ after sampling the posterior using data up to various values of $k_{\rm max}$. The set of parameters that we vary depends on the multipoles that we fit to exactly as in previous sections. Recall that we have the ability to turn FoG off by simply setting the damping factor to unity (blue points); when we do so, constraints are unbiased, as expected given that the input parameters are chosen to lie within one standard deviation of the central value of the prior appropriate for $k_{\rm max}=0.25\,h \rm{Mpc}^{-1}$; moreover, with such large effective volumes, the impact of the prior should be limited anyway. Once we turn FoG on (orange points) cosmological constraints become biased, more so the higher $k_{\rm max}$. The deviation appears to be greater when the hexadecapole is included --- even though in this case we vary three additional parameters ($b_3, \alpha_4$ and $\rm{SN}_4$) ---, exceeding the 5\% level for $\omega_c$ and $A_s^{1/2}$ by $k_{\rm max}=0.2\,h\,\rm{Mpc}^{-1}$. This suggests that the hexadecapole is more sensitive to redshift-space non-linearities, as has been pointed out before in the literature (e.g. Ref.~\cite{maus_analysis_2024}). However, biases of several percent appear already by $k_{\rm max}=0.15\,h\,\rm{Mpc}^{-1}$ even when fitting to only the monopole and quadrupole. Note that including the additional two-loop-order counterterm (empty points) does not improve the situation significantly.

\begin{figure}
    \centering
\includegraphics[scale=0.74]{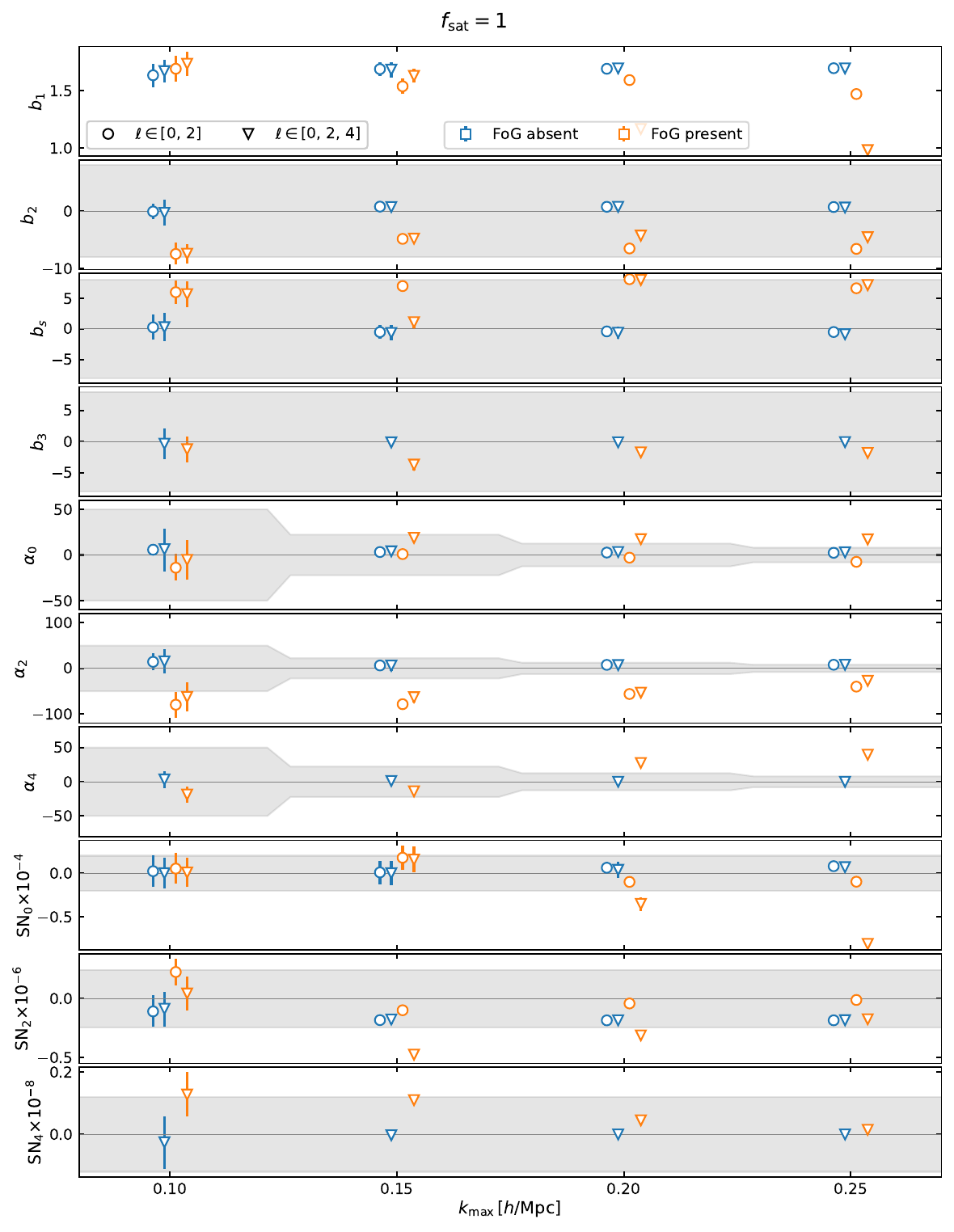}
    \caption{Marginalized posterior mean and 68\% credible intervals on nuisance parameters as a function of $k_{\rm max}$ associated with the inferences in figure~\ref{fig:constraints_sats_only}, which consider composite, synthetic data vectors with $f_{\rm sat}=1$. We overlay also the 68\% confidence interval from the prior (shaded regions; the prior on $b_1$ is uniform so we do not show it). Note that when FoG are present, the counterterm parameters, $\alpha_n$, take on values disfavored by the prior, suggesting a breakdown of perturbativity.}
    \label{fig:cosmo_constr_toymodel_satsonly}
\end{figure}

It is instructive to investigate the direction in which FoG drive the EFT parameters. In figure~\ref{fig:cosmo_constr_toymodel_satsonly}, we show the marginalized constraints on the nuisance parameters corresponding to the inferences above. In addition to the posterior mean and 68\% credible intervals, we overlay also the prior. The reader may recall that the prior on the counterterm parameters was chosen so as to ensure that the EFT contributions remain a small fraction of the linear-theory contribution at $k_{\rm max}$; more specifically, one prior standard deviation away from zero, $\alpha_{n}$ constitutes a 50\% correction to the linear-theory prediction for the $n$-th moment~\cite{maus_analysis_2024}. From figure~\ref{fig:cosmo_constr_toymodel_satsonly}, we learn that when FoG are present, the preferred counterterm parameter values lie in regions that are disfavored by the prior.  $\alpha_2$, in particular, lies more than one standard deviation away from the prior center already at $k_{\rm max}=0.1 \,h \rm{Mpc}^{-1}$ whether or not the hexadecapole is included. This points to a breakdown of perturbativity roughly at the scale where the quadrupole crosses zero. Note, however, that \emph{all} counterterms respond: this is because stochastic velocities introduce a number of counterterms into the theory, all but one of which are fully degenerate with the redshift-space counterterms.

\subsubsection*{Tests on a toy model of typical LRG data}
We now turn to quantifying the impact of FoG non-linearities on a more realistic scenario where satellites coexist with centrals. We will follow the same strategy as above and attempt to fit a `composite', synthetic data vector with the original redshift-space EFT functional form, which, importantly, features no \emph{ad hoc} FoG damping. Once again, we fix the satellite velocity dispersion to a conservative $\sigma_v=7h^{-1}\rm{Mpc}$, but this time we vary the satellite fraction to regulate the importance of FoG. By deviating from the $f_{\rm sat}=1$ of the previous section, we are now also sensitive to the contribution from central-satellite pairs.

\begin{figure}
    \centering
\includegraphics[scale=0.77]{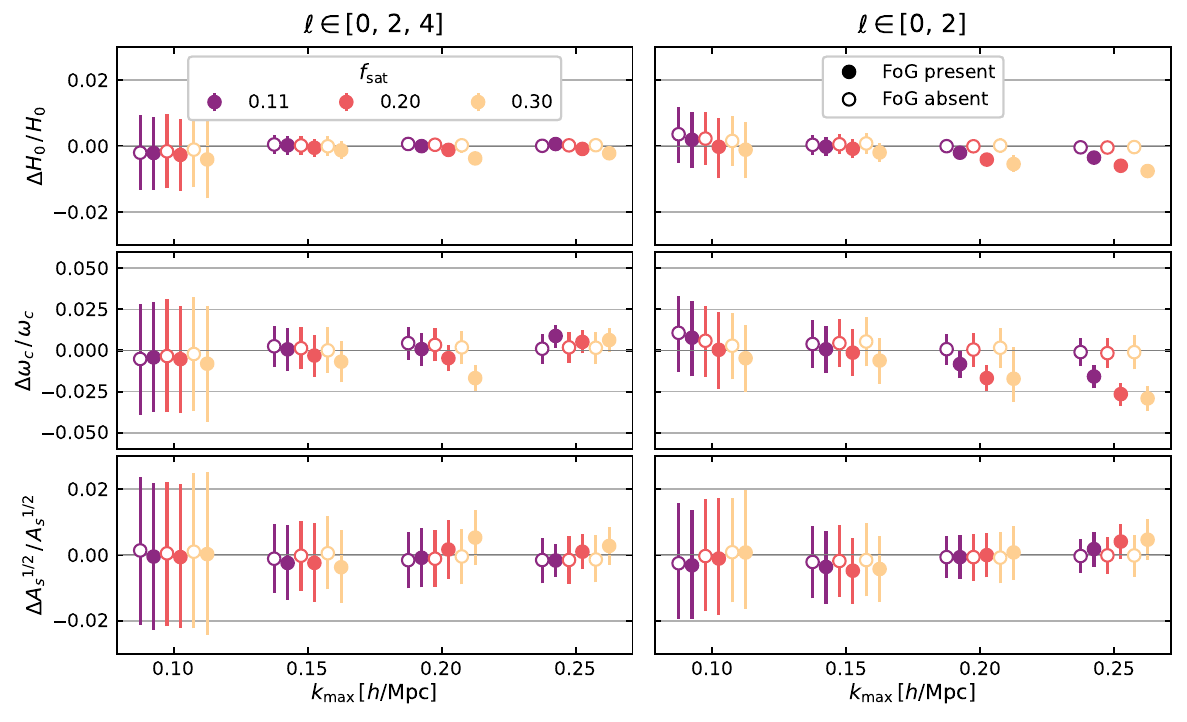}
\caption{Marginalized posterior mean and 68\% credible intervals on cosmological parameters as a function of $k_{\rm max}$, obtained from analyzing a noiseless, synthetic EFT+damping data vector for which $\sigma_v=7h^{-1}\rm{Mpc}$. Different colors correspond to different values of the satellite fraction --- higher values correlate with stronger FoG. When no FoG damping is applied (empty markers) constraints are unbiased; but when a phenomenological damping resembling FoG is applied (filled) the inferred cosmology is biased, the more so the higher the satellite fraction. Circles correspond to fits to the monopole and hexadecapole, while inverted triangles include the hexadecapole as well (in which case we vary also $\alpha_4, \rm{SN}_4$ and $b_3$). Constraints are plotted as a fractional deviation from the true parameter value underlying the simulations.}\label{fig:constraints_tot_toymodel}
\end{figure}

Figure~\ref{fig:constraints_tot_toymodel} shows the marginalized constraints on cosmological parameters as a function of $k_{\rm max}$ for three values of the satellite fraction, $f_{\rm sat}=\{0.11, 0.20, 0.30\}$. The first of these, $f_{\rm sat}=0.11$, is the value measured from the our reference \texttt{Abacus} simulation. The figure shows a steady increase in bias with both $k_{\rm max}$ and $f_{\rm sat}$. We verify that in all cases the bias disappears if the FoG damping factor is replaced with unity (empty markers), indicating that the bias can be attributed to FoG. For reference, the full-shape analysis of all DESI Y1 samples within $\Lambda$CDM constrains $H_0$, $\Omega_m$ and $\sigma_8$ (for which $A_s$ is a good proxy) to $1.1\%, 3.2\%$ and $4\%$, respectively~\cite{collaborationDESI2024VII2024}. This precision will of course improve as more data is gathered and analyzed, increasing the sensitivity to systematic effects such as FoG.

At this point, it is worth noting that the lack of two-loop effects in this toy model makes a direct comparison between figures~\ref{fig:constraints_tot_toymodel} and~\ref{fig:cosmo_constraints_abacus}  unwarranted. Nevertherless, we do note that the trend seen in figure~\ref{fig:cosmo_constraints_abacus} where going from $f_{\rm sat}=0$ to $f_{\rm sat}=0.11$ shifts $\Delta \omega_c$ and $\Delta H_0$ in the negative direction and  $\Delta A_s$ in the positive is reproduced also in figure~\ref{fig:constraints_tot_toymodel} as the satellite fraction is increased.

\begin{figure}
    \centering
\includegraphics[scale=0.74]{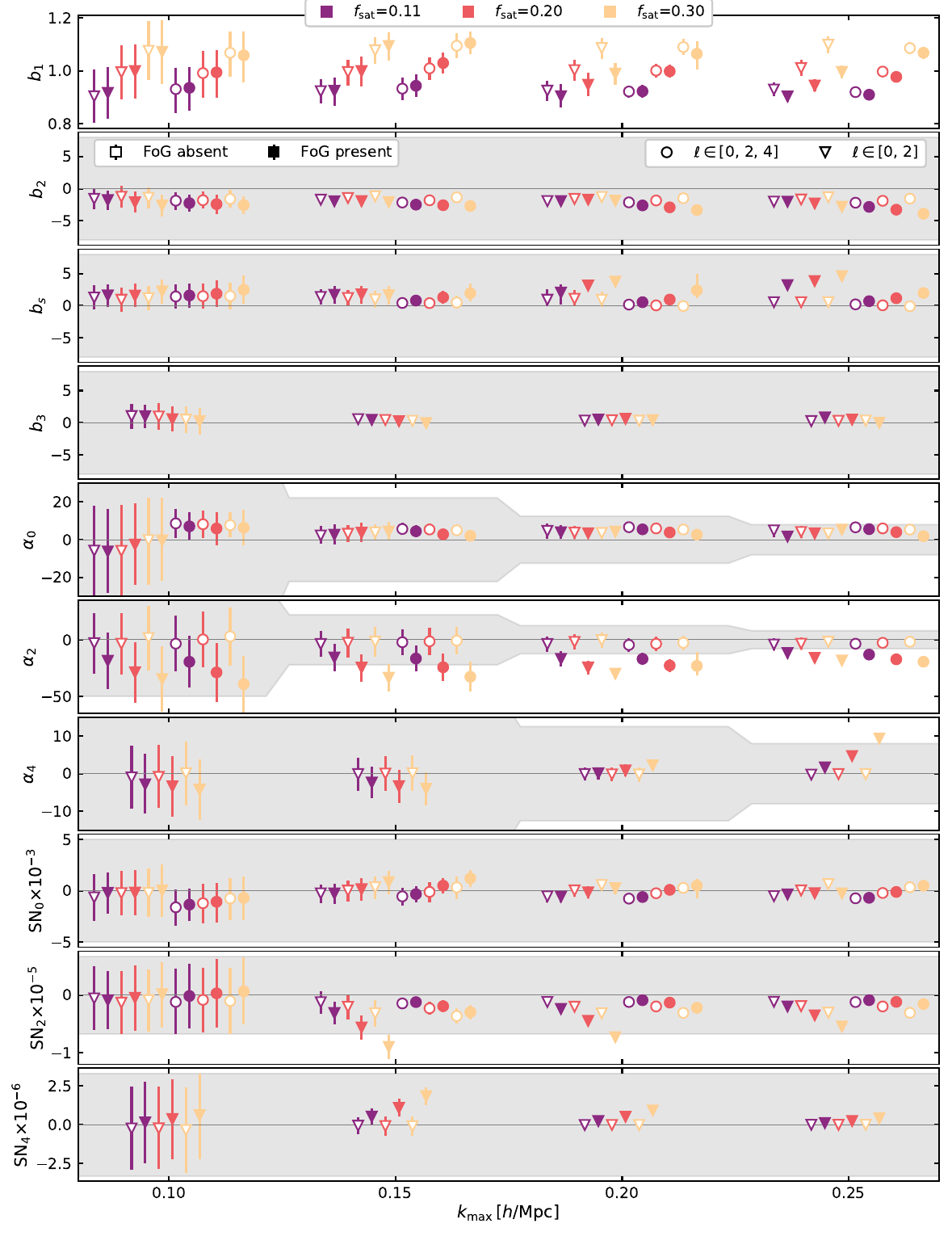}
    \caption{Marginalized posterior mean and 68\% credible intervals on nuisance parameters as a function of $k_{\rm max}$ for the inferences in figure~\ref{fig:constraints_tot_toymodel}. We overlay also the 68\% confidence interval from the prior (shaded regions). The counterterm parameters, $\alpha_n$, absorb contributions from FoG; when they take on values disfavored by the prior, perturbativity is broken.}
    \label{fig:marginals_cosmo_toymodel_tot}
\end{figure}

Once again, it is worth visualizing the constraints on nuisance parameters corresponding to these inferences. These are shown in figure~\ref{fig:marginals_cosmo_toymodel_tot}. The striking growth in $b_1$ with the satellite fraction is due to the mock sample having a higher proportion of satellites, which have higher bias. More interestingly, the $\alpha_n$ counterterms absorb FoG contributions and become large compared to the standard deviation of the prior imposed on them, which suggests that the counterterms are no longer a small correction to the linear theory contribution. This is important for two reasons: first, it tells us that the perturbative expansion is breaking down; and second, in current data analyses, the priors imposed on the counterterms are typically chosen to be informative in order to avoid prior volume effects. For a given prior on $\alpha_n$, mischaracterizing the impact of FoG contributions to this parameter can effectively shrink the prior on contributions from other processes and potentially bias cosmological constraints. 

\section{Practical implementations of FoG mitigation}\label{sec:examples}
Having established that the satellite-induced FoG contributions are non-perturbative already at moderate scales, it follows that the only way to mitigate this nuisance (at least while fitting perturbative models) is to remove from the sample the galaxies most responsible for this effect. In this section, we explore two ways to do this of immediate applicability: simple cuts in galaxy color space, and discarding galaxies in sky regions with bright tSZ emission. 

\subsection{Discarding the reddest galaxies}\label{sec:discarding_reddest}
 
The connection we showed between the zero-crossing of the quadrupole and the prevalence of FoG suggests using this observable as a data-driven and model-independent way to select samples less affected by FoG. One could imagine defining a selection in galaxy color space, measuring the quadrupole, perturbing the selection, referring to the quadrupole again, and iterating this procedure until the tradeoff between minimizing the quadrupole zero-crossing and respecting some other constraints (e.g. the number density of galaxies) is optimized. Though investigating this approach in detail is beyond the scope of this particular paper, we will now explore what can be considered the first step in this iteration: discarding the reddest galaxies in a given sample. It is well known that red galaxies display stronger FoG~\cite{coilDEEP2GalaxyRedshift2008, mohammadVIMOSPublicExtragalactic2018, hangGalaxyMassAssembly2022, mergulhaoEffectiveFieldTheory2023}. This is related to the fact that stochastic velocities originate primarily from galaxies living in the most massive halos, and such high density regions are predominantly populated by redder galaxies, whereas bluer ones tend to avoid such environments~\cite{dresslerGalaxyMorphologyRich1980, madgwick2dFGalaxyRedshift2003}. Indeed,
Ref.~\cite{singh_fundamental_2021} (hereafter SYS21), showed that the zero-crossing of the quadrupole is very sensitive to the color of galaxies (see their figure~15), and, to a lesser extent, also to their luminosity (their figure~14). Concretely, they showed that the power spectrum quadrupole of the reddest 20\% of BOSS LOWZ galaxies crosses zero at $k\approx0.2\,h\,\rm{Mpc}^{-1}$, while the full sample has a crossing at $k\approx0.3\,h\,\rm{Mpc}^{-1}$. This suggests that splitting samples in color and simply dropping the reddest quintile could help mitigate FoG. Since the color selection is `local' (in the sense that it is sensitive to the properties of individual galaxies rather than the large-scale structure of the Universe), this selection should not introduce the kinds of modeling complications that plague alternative approaches to FoG removal.

When trying to study the impact of such color-based selection, we face the difficulty that the \texttt{Abacus} simulations do not mock up the color of galaxies. In spite of this, we will still be able to approximate different color subsamples somewhat roughly by ensuring that they reproduce the power spectra of the color splits reported in SYS21. Let us now explain how this can be done. 

First, following \S\ref{sec:hod_results}, we reweight satellites and centrals in the fiducial BOSS CMASS HOD of Ref.~\cite{yuan_span_2022} (modified to have $f_{\mathrm{ic}}=1$ as described in \S\ref{sec:breakdown_of_Pk}) in order to better match the zero-crossing of the full LOWZ sample as reported in SYS21. We find that a weighting that leads to an effective satellite fraction of $f_{\rm sat}=0.15$ gives a crossing at $k=0.31 h\,\rm{Mpc}^{-1}$, similar to that seen in the data. Similarly, $f_{\rm sat}=0.3$ reproduces the zero-crossing that SYS21 finds at $k=0.21 h\,\rm{Mpc}^{-1}$ for the reddest 20\% of galaxies. Note that in the `full sample' there are $0.15 \bar{N}^{\rm full} = 0.75 \bar{N}^{\rm red}$ satellites, while in the red subsample the number is $0.3 \bar{N}^{\rm red}$ ($\bar{N}$ denotes the number of galaxies in a given sample). This already suggests that in dropping the reddest quintile we would be discarding 40\% of all the satellites, such that the remaining `blue' subsample will have a satellite fraction of $f_{\rm sat}^{\rm blue} = (f_{\rm sat}^{\rm full}\bar{N}^{\rm full} - f_{\rm sat}^{\rm red}\bar{N}^{\rm red} )/\bar{N}^{\rm blue} = 0.1125$, a 25\% reduction relative to the full sample.

Given that we have approximated the power spectra of the full sample and the reddest quintile, we can actually combine these to extract the power spectrum of the blue subsample that remains after dropping the reddest galaxies. More explicitly, let us denote the redshift-space power spectrum of the full sample as $P^{\rm{full}}(\bm{k})$, that of some redder subsample as $P^{\rm{red}}(\bm{k})$, and that of the remaining bluer subsample as $P^{\rm{blue}}(\bm{k})$. It is straightforward to show that
\begin{equation}\label{eqn:Pblue}
    P^{\rm{blue}}(\bm{k}) = \left(\frac{\bar{N}^{\rm{full}}}{\bar{N}^{\rm{blue}}}\right)^2 \left( P^{\rm{full}}(\bm{k}) - \left(\frac{\bar{N}^{\rm{red}}}{\bar{N}^{\rm{full}}}\right)^2  P^{\rm{red}}(\bm{k}) - 2\frac{\bar{N}^{\rm{red}} \bar{N}^{\rm{blue}}}{(\bar{N}^{\rm{full}})^2}  P^{\rm{red-blue}}(\bm{k}) \right)\,,
\end{equation}
Since both this expression and equation~\eqref{eqn:multipoles} --- which we use to extract the power spectrum multipoles --- are linear in $P(\bm{k})$, it follows that the relation also holds for the multipoles. 

\begin{figure}
    \centering
\includegraphics[scale=0.74]{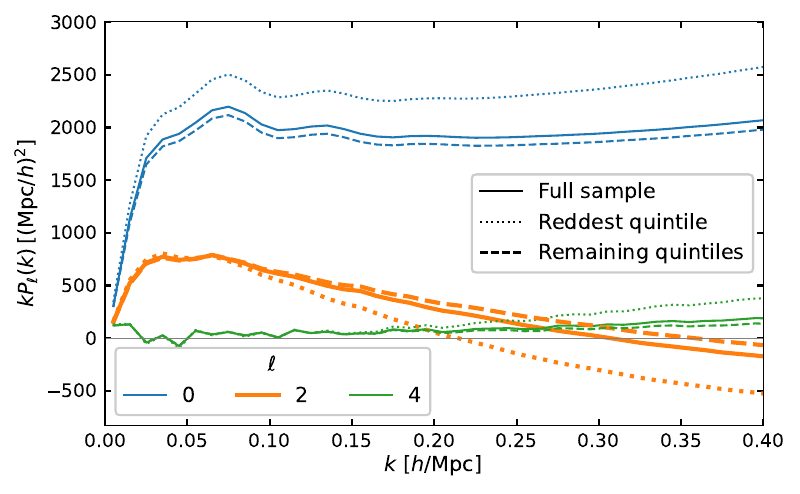}
\caption{Power spectrum multipoles approximating the effect of removing the reddest 20\% of galaxies from a BOSS-LOWZ-like sample. (Unlike in other plots in this work, we do not subtract shot noise from the monopole.) By reweighting galaxies to have an effective $f_{\rm sat}=0.3$ ($f_{\rm sat}=0.15$), we mock up a sample with a zero-crossing of the quadrupole at $k=0.21$ ($k=0.31$) resembling that of the reddest quintile (full sample) of BOSS LOWZ~\cite{singh_fundamental_2021}. From this we can predict the spectra of the remaining 80\% of galaxies after the reddest quintile is discarded, which has a zero-crossing at a higher $k=0.36$. This suggests that the remaining sample has a significantly smaller scale of FoG non-linearity.}
\label{fig:Pk_after_color_split}
\end{figure}

All of these spectra are shown in figure~\ref{fig:Pk_after_color_split}. We learn that discarding the reddest 20\% of galaxies shifts the zero-crossing of the quadrupole from $k=0.31 h\,\rm{Mpc}^{-1}$ to $k=0.36 h\,\rm{Mpc}^{-1}$. Taking figures~\ref{fig:fsat_vs_kcrossing} and~\ref{fig:multipole_alpha_dependence} at face value, this 16\% shift in the zero-crossing scale could be roughly equivalent to lowering the satellite fraction by 25\% (as we already discussed above) or the satellite velocity dispersion by 28\%. 

\begin{figure}
    \centering
\includegraphics[scale=0.77]{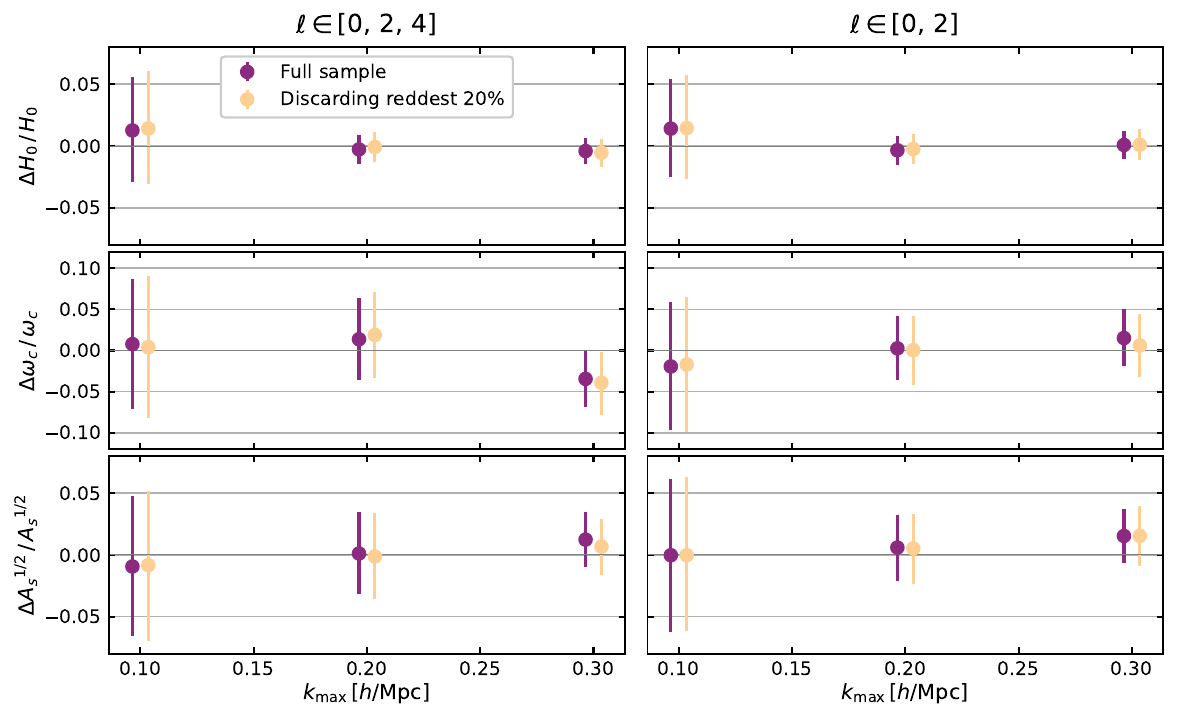}
\caption{Marginalized posterior mean and 68\% credible intervals on cosmological parameters as a function of $k_{\rm max}$ before (purple) and after (yellow) reweighting galaxies in a way that mimics discarding the reddest quintile of mock galaxies resembling the BOSS LOWZ selection.}\label{fig:cosmo_constr_colorselect}
\end{figure}

The question of whether this translates to improved cosmological constraints when fitting an EFTofLSS model is addressed in figure~\ref{fig:cosmo_constr_colorselect}, where we compare parameter recovery before and after discarding the reddest 20\% of simulated galaxies in the approximate way described above. We find marginal improvements in the accuracy of the recovered parameters; as explained above, the limited nature of the gains is likely due to the preponderance of other two-loop effects that cause an earlier breakdown of the modeling.

\subsection{Masking galaxy clusters and groups}\label{sec:masking_clusters}
In \S\ref{sec:HOD_properties}, we showed that FoG are sourced primarily by galaxies living in the most massive halos, where deep gravitational potential wells source large virial velocities. Before quantifying the extent to which removing these galaxies can improve cosmological constraints from the full-shape of the galaxy power spectrum, let us briefly comment on what ancillary data would enable such a selection.

Massive galaxy clusters produce strong X-ray emission and Sunyaev-Zel'dovich (tSZ) distortions of the cosmic microwave background frequency spectrum. tSZ and X-ray intensity are tightly correlated with halo mass, so masking on these observables --- for example, in regions with SNR of 5 or higher --- could potentially remove the galaxies with the strongest peculiar velocities\footnote{There is likely a connection between what we propose to do and density-split statistics or marked power spectra, though we claim that our selection is more local in nature as it is guided by halo mass. Our approach will thus produce populations with clustering properties that are easier to model. The modeling will also be made easier by the fact that we typically identify --- and discard --- a very small number of objects.}. The selection and removal of these galaxies would thus be very clean and depend only on halo mass and crucially \emph{not} on galaxy overdensity or velocity\footnote{This is due in large part to the fact that structures that can be detected on an individual basis in current or upcoming Compton $y$-maps are all in the one-halo regime, with two-halo contributions to the tSZ effect only becoming evident after stacking over very many objects (the impact of two-halo contributions to the $y$-maps could be checked on a case-by-case basis using hydrodynamical simulations). Moreover, these objects are typically identified only after high-pass filtering the Compton $y$-maps to eliminate any long-wavelength modes tracing the cosmic web. This means that the key discerning property in this selection is halo mass. The resulting galaxy sample will simply have a different (smaller) galaxy bias than the initial population; bias can of course be marginalized over in the standard ways, at least in the limit applicable here where only a relatively small fraction of the galaxies are masked.}. (Below, we will in fact verify that thresholding on halo mass does not introduce any noticeable biases on parameter inference; much on the contrary, it helps recover more accurate cosmological parameters.) As an example, suppose we are to analyze a galaxy redshift survey that fully overlaps with a complementary CMB survey. The tSZ signal being independent of redshift, we could in principle have a complete census of all the clusters above a certain mass threshold, and discard galaxies living within them\footnote{Future work will be required to understand the impact of interloping gas, the correlation between window and density field (see e.g.~\cite{lemboCMBLensingReconstruction2022, surraoAccurateEstimationAngular2023} for related investigations), and the precision of available cluster redshifts, though we expect these to be relatively minor issues.}. In its nominal configuration, SO will detect 16,000 tSZ clusters, promising to get near the $M_{\rm{halo}}\sim 10^{14}\,M_{\odot}$ mark~\cite{the_simons_observatory_collaboration_simons_2019, zubeldia_cosmocnc_2024}. This will improve further with CMB-S4, which is expected to detect between 45,000 and 140,000 clusters, depending on the final design choice~\cite{abazajian_cmb-s4_2016}. While CMB measurements are the most promising way to detect clusters at high redshift, X-ray measurements from eRosita~\cite{merloniEROSITAScienceBook2012} promise to deliver a highly complete catalog of clusters at $z<0.5$~\cite{abazajian_cmb-s4_2016}, with masses well below $M_{\rm{halo}}\sim 10^{14}\,M_{\odot}$.

\begin{figure}
    \centering
    \includegraphics[scale=0.74]{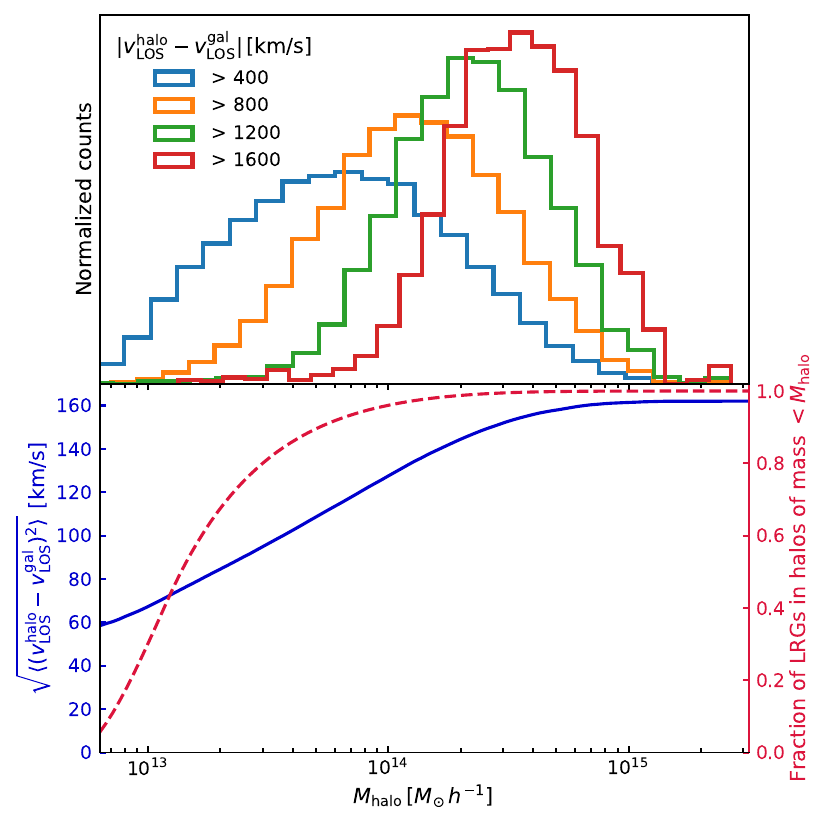}
    \caption{Mitigating fingers-of-God in a sample of LRGs by masking galaxy clusters. Figure~\ref{fig:masking_massive_halos} showed that excising the most massive halos shifts the zero-crossing of the quadrupole --- and thus the scale of FoG non-linearity --- to higher $k$. The reason, illustrated in the top panel of this figure, is that the largest differences between the radial velocity of galaxies and that of their host halo are found in the most massive clusters and groups --- the histograms (shown here for a single \texttt{Abacus} realization) correspond to the 3.3\%, 0.08\%, 0.02\% and 0.005\% of galaxies with the highest radial velocity residual, yielding errors in the inferred redshift-space position of 4, 8, 12 or 16\,$h^{-1}$\,Mpc, respectively. The bottom panel shows that the RMS velocity residual (blue, solid) receives sizable contributions from very massive halos where relatively few LRGs live (in red, dashed, we show the fraction of LRGs that live in halos of mass smaller than $M_{\rm{halo}}$).}\label{fig:masking_tsz}
\end{figure}

Figure~\ref{fig:masking_massive_halos} shows variations of the power spectrum quadrupole after removing the most massive clusters from the CMASS-inspired mock catalog described in \S\ref{sec:abacus}. We can see that removing the most massive clusters shifts the zero-crossing to higher $k$, thus helping mitigate FoG. The reason for this can be understood from the top panel of figure~\ref{fig:masking_tsz}, where we histogram the difference between the line-of-sight velocity of galaxies and that of their host halo as a way to isolate galaxy thermal velocities. This confirms our intuition that the highest stochastic velocities --- those from which we can expect to be able to extract little cosmological information --- are found in the most massive halos.

In this HOD model a mere 0.1\% of galaxies have velocity residuals in excess of around 1000\,km\,s$^{-1}$, leading to radial position errors of around 10\,Mpc$\,h^{-1}$. These galaxies live primarily in halos heavier than $10^{14}\,\mathrm{M}_{\odot}\,h^{-1}$ and can therefore be easily identified with upcoming CMB and X-ray observations and removed. In addition to considering these worst outliers --- which are expected to have an outsize effect on the power spectrum --- we can also paint a richer picture by looking at how the root-mean-squared (RMS) velocity residual and the number of galaxies scale with halo mass (bottom panel\footnote{It is instructive to read our figure~\ref{fig:masking_tsz} alongside figure~3 of
Ref.~\cite{schmittfull_modeling_2021}, which quantifies the velocity residual between similar mock galaxies and a field-level, LPT-based prediction of their velocity correct to linear order in the displacements. Our findings are qualitatively consistent, with any quantitative differences likely due to differences in the samples and the fact that their velocity model is truncated at leading order in the displacements (hence why their total residual is larger).}). We learn that an experiment that can identify all halos above $10^{14}\,M_{\odot} \,h^{-1}$ (not far from expectations for SO) will be able to reduce the RMS velocity residual by 20\% while removing only 4\% of the galaxies.

The next few percent of galaxies with the highest line-of-sight velocity residuals populate halos of a wide range of masses in the cluster and group scale, with some nearing $10^{13}\,\mathrm{M}_{\odot}\,h^{-1}$. Despite the fact that virial velocity is strongly correlated with halo mass, the abundance of very massive clusters is exponentially suppressed at the tail of the halo mass function, so that overall only a small fraction of the galaxies with residuals of a few Mpc$\,h^{-1}$ live in the most massive halos. In general, above approximately $10^{13.5}\,M_{\odot} \,h^{-1}$, the slopes of the RMS velocity residual and galaxy fraction are such that it is favorable to remove the halos, but below this threshold the marginal cost of reducing the RMS is discarding a relatively larger number of galaxies\footnote{For $10^{13} < M_{\rm halo}\, [M_{\odot}\,h^{-1}] < 10^{14}$, the RMS velocity residual becomes a power law, roughly $\rm{RMS} \propto ( M_{\rm halo}^{\rm max})^{0.3}$.}: for example, it would require removing 50\% of the LRGs in order to halve the RMS velocity error. Alternative approaches --- including the one developed in this work that makes use of the zero-crossing of the power spectrum quadrupole --- may be more effective when it comes to removing problematic galaxies in these less massive environments.

\begin{figure}
    \centering
\includegraphics[scale=0.77]{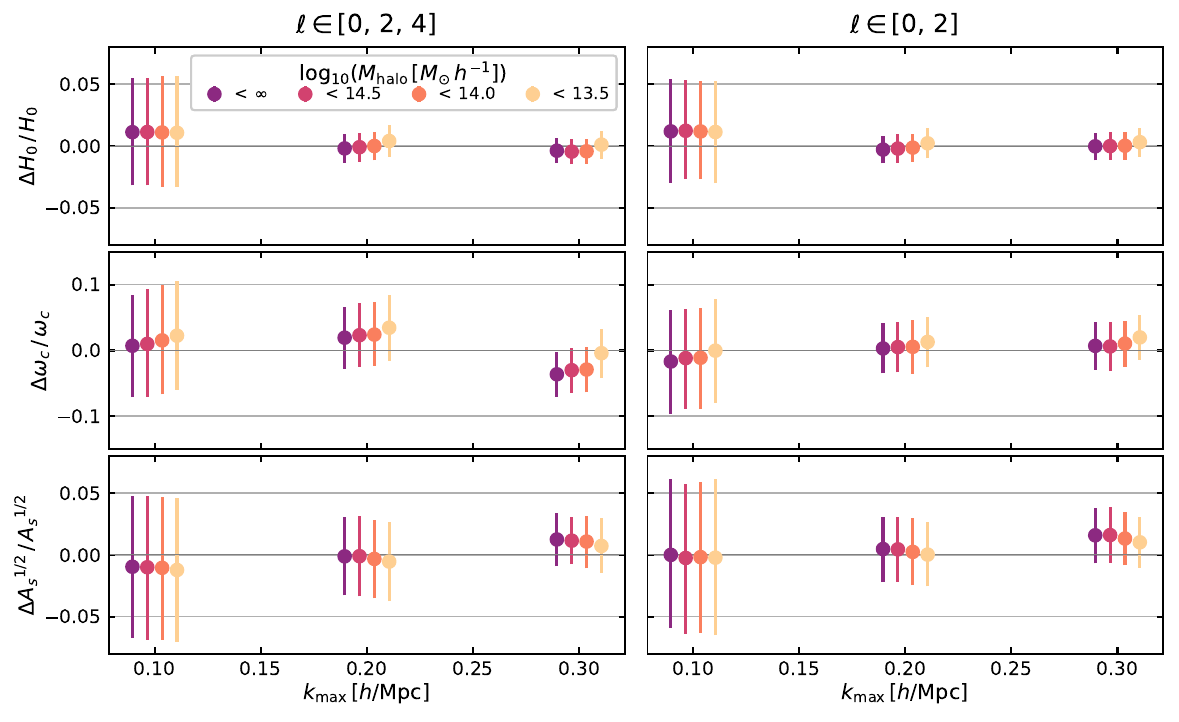}
\caption{Marginalized posterior mean and 68\% credible intervals on cosmological parameters as a function of $k_{\rm max}$ from analyzing simulations with various maximum halo mass cuts.}\label{fig:cosmo_constr_discarding_clusters}
\end{figure}

In figure~\ref{fig:cosmo_constr_discarding_clusters}, we show cosmological constraints resulting from analyzing the \texttt{Abacus} simulations with the different maximum-halo-mass cuts discussed above. To facilitate comparison, we use the same prior for all cases: that of table~\ref{tab:priors} with $f_{\rm sat}=0.11$ and $\sigma_v=7h^{-1}\rm{Mpc}$. Despite the obvious impact of two-loop corrections, which imprint residuals that grow with $k_{\rm max}$, it is readily apparent that, at a given $k_{\rm max}$, lowering the maximum halo mass tends to shift the central value of constraints towards the truth. Notice, in particular, that imposing $M_{\rm halo} < 10^{13.5} [M_{\odot}\,h^{-1}]$ gives rise to constraints that are consistent with the truth within 1-$\sigma$ of the 25-box uncertainty when fitting $\ell\in[0,2,4]$ up to $k_{\rm max}=0.3$.\footnote{In all the constraints quoted in this section, the $\alpha_{(k\mu)^4}$ counterterm is fixed to zero.}
\section{Discussion and conclusions}\label{sec:discussion}
The peculiar velocity of galaxies induces redshift-space distortions of their clustering pattern. This can be both a blessing and a curse: while distortions sourced by the large-scale velocity field contain valuable information about the growth of cosmic structures and the matter content of the Universe, the contribution associated with local dynamics --- the Fingers-of-God pattern --- poses a great challenge to modeling efforts as it is highly non-linear.

In \S\ref{sec:eft_tests}, we showed explicitly that cosmological constraints derived from fitting EFTofLSS models to redshift-space power spectrum multipoles are suceptible to bias when FoG are present. The reason for this was explained in \S\ref{sec:intro_to_RSD}, where we used a simplified model to argue that the clustering of central-satellite and satellite-satellite pairs is expected to escape perturbative control on scales where their power spectrum quadrupole is negative. In \S\ref{sec:breakdown_of_Pk}, we took this insight further: we simulated a typical galaxy sample by populating the \texttt{Abacus} $N$-body mocks with galaxies resembling BOSS CMASS LRGs using an HOD, and found that the satellite-satellite contribution crosses this threshold around $k\approx 0.1\,h\,\rm{Mpc}^{-1}$, whereas for central-satellite pairs this is closer to $k\approx 0.2\,h\,\rm{Mpc}^{-1}$. From this, we argued that, already by $k\approx 0.2\,h\,\rm{Mpc}^{-1}$, there are fully non-perturbative contributions to the power spectrum multipoles at the $\mathcal{O}(10\%)$ level per mode, growing rapidly with $k$. This suggests that beyond those scales, FoG should be considered a source of unmodelable systematic error that is better removed\footnote{Alternative models have recently been developed which pursue a non-perturbative treatment of the FoG (e.g.~\cite{eggemeierBoostingGalaxyClustering2025}). If these are shown to describe FoG in realistic ways, they will constitute an interesting complement to the removal we advocate for.}.

To this end, we set out to identify indicators of FoG abundance that could be used to select cleaner samples. The crucial requirement on these indicators is that they be data-driven and that they make no explicit reference to the redshift-space galaxy density field itself. The latter point is key to avoid introducing new issues such as velocity bias, which affect FoG compression techniques proposed previously in the literature (e.g.~\cite{tegmark_cosmological_2006, reid_luminous_2009}). Ideally, the new techniques will only depend on local properties of the galaxy or its immediate environment, and not the large-scale structure it is embedded in.

As mentioned already, we identify the zero-crossing of the power spectrum quadrupole as a robust indicator of FoG abundance in a given sample, showing in \S\ref{sec:HOD_properties} that it is inversely proportional to the satellite fraction (cf. figure~\ref{fig:fsat_vs_kcrossing}) and the satellite velocity dispersion (cf. figure~\ref{fig:multipole_alpha_dependence}). One could then imagine selecting samples of galaxies with fewer FoG by iteratively exploring different regions of the color-color plane while referring to the quadrupole at every stage, striving to maximize the zero-crossing while respecting any other constraints pertinent to the desired targets. While we intend to explore this in future work, a more immediate implementation would be to discard the reddest galaxies in the sample (see also Refs~\cite{coilDEEP2GalaxyRedshift2008, mohammadVIMOSPublicExtragalactic2018, hangGalaxyMassAssembly2022, mergulhaoEffectiveFieldTheory2023}): in \S\ref{sec:examples}, we show that removing the reddest 20\% of CMASS galaxies could shift the zero-crossing from $k=0.31\, h\,\rm{Mpc}^{-1}$ to $k=0.36\, h\,\rm{Mpc}^{-1}$, roughly equivalent (according to figure~\ref{fig:fsat_vs_kcrossing}) to lowering the satellite fraction by 25\%, or the satellite velocity dispersion by 28\% (from figure~\ref{fig:multipole_alpha_dependence}).

We also find that many of the galaxies producing the Fingers-of-God live in massive halos which can be identified and excised from the maps using ancillary data. The most promising observable to this end is probably the tSZ spectral distortion of the CMB, since its distinct frequency spectrum allows it to be cleanly disentangled from other sources of microwave emission, and the signal is redshift-independent and tightly correlated with halo mass. At low redshift, X-ray observations should also be a powerful complement. We show that an experiment that can identify and remove all clusters above $10^{14}\,M_{\odot} \,h^{-1}$ --- which SO and/or CMB-S4 should be able to do --- will get to reduce the RMS velocity residual between galaxies and halos by 20\% while removing only 4\% of the galaxies.

At low redshift, mitigating FoG does not immediately extend the regime of validity of one-loop EFTofLSS models in a significant way. This is most likely due to non-linear velocities or two-loop effects different from FoG becoming relevant already on moderate scales. There is abundant evidence of this in the literature: figure 6 of Ref.~\cite{ivanov_cosmological_2021} shows that the accuracy in the determination of cosmological parameters when fitting one-loop theory spectra improves only moderately as the satellite fraction in simulations of emission-line galaxies (from Ref.~\cite{avila_completed_2020}) is lowered; even with $f_{\rm sat}=0$, the inferred $\sigma_8$ becomes biased at the percent-level already at $k_{\rm{max}}\approx 0.24\,h\,\rm{Mpc}^{-1}$. Similarly, Ref.~\cite{chen_consistent_2020} fit the power spectrum quadrupole of halos in redshift space out to $k\approx 0.25 \, h\,\rm{Mpc}^{-1}$ and obtained fits that were at best 2-3\% accurate\footnote{In contrast, they were able to fit most wedges with percent-level precision. This points to high-$\mu$ non-linearities, which are mixed together with low-$\mu$ contributions in the ($\ell\neq 0$) multipoles~\cite{chen_consistent_2020}.}. Analogous conclusions can be drawn from the field-level analysis of Ref.~\cite{schmittfull_modeling_2021}, who found that the $k$-reach of the model could not be extended significantly even after the galaxies with the highest thermal velocities were removed\footnote{Private communication from Misha Ivanov.}. More explicitly, Ref.~\cite{taule_two-loop_2023} calculated the two-loop EFTofLSS model for matter in redshift space and compared it to $N$-body simulations, demonstrating that two-loop corrections become significant already at $k\approx 0.15\,h\,\rm{Mpc}^{-1}$ for $z=0.5$, and are essential to fit the data beyond $k\approx 0.20\,h\,\rm{Mpc}^{-1}$ given the error bars of the simulations. If these higher-order corrections are indeed within reach of perturbation theory --- or if an alternative model that is less affected by them is developed --- then removing the non-perturbative FoG as proposed in this work could extend the applicability of the models to smaller scales, improving constraining power. Having access to more constraining data could in turn alleviate issues to do with prior volume effects, which are currently a significant nuisance in Bayesian analyses of galaxy clustering using the EFTofLSS~\cite{maus_comparison_2023, simon_consistency_2023, holm_bayesian_2023}. Removing FoG might also help mitigate the evolution of EFT parameters with redshift, which is a challenge for the simulation-based calibration of EFT priors~\cite{ivanovFullshapeAnalysisSimulationbased2024, zhangHODinformedPriorEFTbased2024}.

In addition to improving the accuracy of `full-shape' analyses of galaxy clustering, removing FoG offers advantages across a wide range of applications in cosmology. An example is reconstructing the large-scale displacement or velocity fields from the redshift-space positions of galaxies. 
The velocities associated with FoG are sourced locally and are thus largely uncorrelated with these large-scale fields. Quantitatively, we can see from the right panel of figure~\ref{fig:masking_tsz} that approximately 1 in every 2,000 CMASS galaxies has peculiar velocity of $\mathcal{O}(1000\,\rm{km}\,\rm{s}^{-1})$. Interpreting the associated redshifts as distance will induce errors of $\mathcal{O}(10\,\rm{Mpc})$ along the line of sight. This can interfere with the reconstruction of the radial BAO feature, which receives 30\% of its signal from $k>0.1\,h\,\rm{Mpc}^{-1}$, corresponding to scales smaller than $\mathcal{O}(10\,\rm{Mpc})$~\cite{eisenstein_improving_2007}. It can potentially also be an issue for reconstructions of the large-scale velocity field, and removing problematic satellites can potentially help improve reconstructions by accessing small-scale information that is currently discarded precisely because of FoG; moreover, discarding these problematic galaxies is not expected to degrade the efficiency of a reconstruction process that is typically saturated in terms of the number of galaxies available~\cite{guachalla_velocity_2024}. In addition to better reconstructions, reducing the satellite fraction can be useful to avoid issues to do with miscentering of gas profiles when stacking on kSZ or tSZ.

Finally, we also note that the zero-crossing of the quadrupole can be used to gauge whether simulations accurately reproduce the non-linearities of the data. As an example, it is worth noting that the recent `PT challenge'~\cite{nishimichi_blinded_2020} relied on mocks that were intended to mimic BOSS CMASS clustering, but display a zero-crossing at $k\approx0.52\,h\,\rm{Mpc}^{-1}$, well beyond the value of $k\approx0.32\textup{--}0.35\,h\,\rm{Mpc}^{-1}$ seen in the data~\cite{yu_rsd_2023}. This suggests that the mocks underrepresent FoG non-linearities, and remind us that the range of validity of theoretical predictions ought to be determined by comparison to many different HOD models, among other robustness tests.

\section*{Acknowledgments}
We are grateful to a number of people for useful conversations that enriched this work: Jamie Sullivan, Boryanah Hadzhiyska, Petter Taule, Edmond Chaussidon, Noah Weaverdyck, Pat McDonald, Martin White, Zvonimir Vlah, Shi-Fan Chen, Emanuele Castorina, Misha Ivanov, Haruki Ebina, Emmanuel Schaan, Henrique Rubira, Hong Guo and especially Mark Maus for his invaluable help with the theory and practice of full-shape modelling. We also thank Martin White and Misha Rashkovetskyi for useful comments on our manuscript.
S.F. is supported by Lawrence Berkeley National Laboratory and the Director, Office of Science, Office of High Energy Physics of the U.S. Department of Energy under Contract No.\ DE-AC02-05CH11231.

This work was carried out on the territory of xučyun (Huichin), the ancestral and unceded land of the Chochenyo speaking Ohlone people, the successors of the sovereign Verona Band of Alameda County.

\appendix

\section{Technical details of our fits to Abacus}\label{appendix:abacus_appendix}

\subsection{Power spectrum measurements}\label{appendix:pk_measurements}
The periodic boxes we analyze have a trivial, homogeneous window function. The power spectrum multipoles can therefore be extracted without recourse to `randoms' that define the local mean density in the absence of clustering (see, e.g., Ref.~\cite{nishimichi_blinded_2020}). We do so using \texttt{PyPower}\footnote{\url{https://github.com/cosmodesi/pypower/tree/main}}, cross-checking our results against the functionality built into \texttt{AbacusUtils}\footnote{\url{https://abacusutils.readthedocs.io/en/latest/index.html}} as well. These codes implement the FFT-based estimator of~\cite{hand_optimal_2017}, which is optimal in this case as the line of sight is plane-parallel by construction (it points along one of the axes of the box). The estimator includes interlacing and a Triangular Grid Assignment (TSC) interpolation scheme which guarantees negligible errors due to aliasing when analyzing scales up to the Nyquist frequency~\cite{sefusatti_accurate_2016}, which in our case is $k_{\rm Nyq}=0.8\,h$\,Mpc$^{-1}$, or half the sampling frequency of the grid. Finally, we bin the measured power spectra into 50 regularly-spaced bins between $0<k\,[h\,\rm{Mpc}^{-1}]<0.5$ giving a grid spacing of $\Delta k = 0.01\,h\,\rm{Mpc}^{-1}$.

In what follows, we will show power spectrum measurements averaged among the 25 \texttt{Abacus} boxes. Error bars on these measurements are drawn from simulation-based covariance matrices, with details provided in appendix~\ref{appendix:mock_covariances}.

\subsection{Rebinning}\label{appendix:discreteness}
We incorporate discreteness effects induced by our periodic box -- see, e.g., \S5.1 of \cite{beutler_clustering_2017} -- using infrastructure built into \texttt{PyPower}. This gives us a matrix that we can apply to our model prediction to effectively `rebin' it while accounting for the incomplete sampling of $(k, \mu)$, particularly at low $k$.

\subsection{Covariances from simulations}\label{appendix:mock_covariances}
We consider only the disconnected part of the covariance. The main reason for this is that every time we vary our HOD choices, the clustering changes, and a different covariance matrix is technically required. While it is easy to re-calculate the disconnected piece, doing so for the connected part is much more costly: it would require on the order of thousands of simulations of a similar volume to what we are considering, each processed with the same HOD. This would constitute a computational challenge far exceeding the scope of this work. Though a connected contribution is indeed expected due to non-linear evolution, its impact on constraints is small. Ref.~\cite{yu_rsd_2023} showed that adding the connected components degraded constraints from the BOSS monopole and quadrupole by only 10-20\% when fitting up to $k_{\rm max}\approx 0.4 \, h /\rm{Mpc}$; the effect is expected to be even smaller for the much larger effective volumes we consider in this analysis (note that the \texttt{PTchallenge}~\cite{nishimichi_blinded_2020} also ignored the connected contribution). Other aspects of the covariance matrix computation also simplify greatly when working with periodic boxes such as ours: the window function is trivial, and there is no super-sample covariance, for example.

The calculation is as follows. If $\hat{P}_{\ell}(k_i)$ are the binned, measured power spectrum multipoles (including shot noise), the covariance of the measured power spectrum multipoles can be calculated as (e.g.,~\cite{nishimichi_blinded_2020})
\begin{align} \label{eq:covmat}
\mathrm{Cov}_{ij}^{\ell \ell'} &= \left\langle\left(\hat{P}_\ell(k_i)-\langle\hat{P}_\ell(k_i)\rangle\right)\left(\hat{P}_{\ell'}(k_j)-\langle\hat{P}_{\ell'}(k_j)\rangle\right)\right\rangle,\nonumber\\
&= \delta_{ij}^\mathrm{K}\frac{(2\ell+1)(2\ell'+1)}{N_i^2} \sum_{\bm{k} \in \mathrm{bin}\,i}\mathcal{L}_\ell(\mu_{\bm{k}})\mathcal{L}_{\ell'}(\mu_{\bm{k}})
\left[P(\bm{k})+P_\mathrm{shot}\right]^2\,,
\end{align}
where $P(\bm{k})$ is the signal component of the redshift-space power spectrum, $P_{\rm shot}$ is the shot noise, and $N_i$ is the number of discrete modes in each $k$-bin. It is worth emphasizing that the sum above runs over all discrete wavenumbers on our measurement grid that fall within each of the $k$-bins. To mitigate Monte Carlo noise, we follow~\cite{nishimichi_blinded_2020} and first average the power spectrum measurement over 10 bins (or `wedges') in $0\leq|\mu_{\bm{k}}|\leq 1$, and over all 25 simulated realizations. On the other hand, the Legendre polynomials are evaluated exactly at each grid element. By matching exactly the grid properties and $k$-binning scheme used to obtain the data vector, we naturally take into account the discreteness effects associated with measuring $P(\bm{k})$ on a grid. Note that this formulation of the covariance allows for non-zero covariance between different multipoles at the same $k$-bin while ensuring that it is zero across different $k$-bins.

\begin{figure}
    \centering
\includegraphics[scale=0.9]{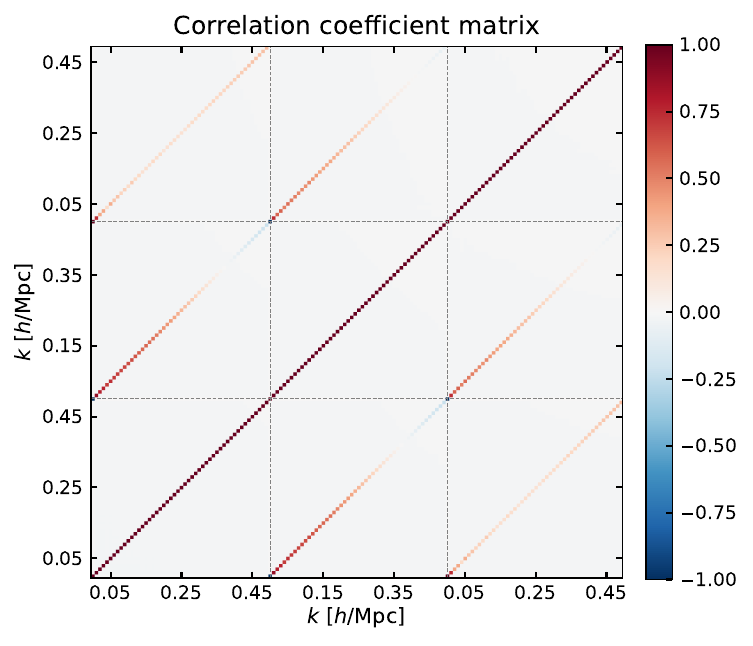}
    \caption{Correlation coefficient of the covariance matrix of power spectrum multipoles for the standard CMASS-like scenario of \S\ref{sec:breakdown_of_Pk}. The correlation coefficient is defined as $\rm{Corr}(\mathcal{O},\mathcal{O}')=\rm{Cov}(\mathcal{O},\mathcal{O}')/[\rm{Cov}(\mathcal{O},\mathcal{O})\rm{Cov}(\mathcal{O}',\mathcal{O}')]^{1/2}$, where $\mathcal{O}\in \{P_0(k), P_2(k), P_4(k)\}$ and $\mathcal{O}'\in \{P_0(k'), P_2(k'), P_4(k')\}$. Blocks are ordered from bottom to top and left to right.}
    \label{fig:corrcoeff}
\end{figure}

Finally, we scale the covariance by a factor commensurate with the effective volume from which our data vector is obtained --- when averaging the measurement over 25 realizations, this means rescaling the covariance by $1/25$. As an example, figure~\ref{fig:corrcoeff} shows the correlation matrix (calculated from the covariance matrix) of the power spectrum multipoles measured from the simulation described in \S\ref{sec:breakdown_of_Pk} (the `full sample', specifically) which is intended to resemble a selection of BOSS CMASS LRGs.

\subsection[AP]{Alcock-Paczynsky distortions}\label{sec:AP}
At the most fundamental level, galaxy surveys observe redshifts and angles. On the other hand, theoretical models provide predictions in terms of spatial distances or scales. To make these two amenable to comparison, observed angles and redshifts must be converted to distances, and for this a fiducial cosmological model must be assumed. Importantly, deviations of this fiducial model from the truth will introduce anisotropy in the clustering irrespective of RSD. These are the famous Alcock-Paczynsky (AP) distortions, which are a valuable cosmological probe in their own right~\cite{alcock_evolution_1979}.

How do they affect our analysis of the Abacus mocks? The only difference with analysis of real data is that here the fiducial cosmology used to convert between redshift and angles into distances is actually the true cosmology. Note that this does not mean that one can ignore AP distortions: they are readily introduced as soon as the mock data is compared to any model different from the truth, as the latter will provide a different translation between angles and redshifts to distances. To remain agnostic about AP and the underlying cosmology, an observer should adjust the wavenumbers and angles in their model as (e.g.~\cite{chudaykin_non-linear_2020})
\begin{equation}
    k \rightarrow k' \equiv k \left[ \left(\frac{H_{\rm true}}{H_{\rm fid}}\right)^2 \mu^2 + \left(\frac{D_{A,\rm fid}}{D_{A,\rm true}}\right)^2 (1-\mu^2)\right]^{1/2}\,,
\end{equation}
\begin{equation}
    \mu \rightarrow \mu' \equiv \mu \left(\frac{H_{\rm true}}{H_{\rm fid}}\right) \left[ \left(\frac{H_{\rm true}}{H_{\rm fid}}\right)^2 \mu^2 + \left(\frac{D_{A,\rm fid}}{D_{A,\rm true}}\right)^2 (1-\mu^2)\right]^{-1/2}\,,
\end{equation}
where $D_{A}$ is the angular diameter distance and $H$ is the Hubble parameter. Here, the subscript `true' refers to parameter values in the proposed model, while the `fid' subscript refers to the assumed model used when constructing maps from catalogs.

\subsection{Emulator}\label{appendix:emulator}
The \texttt{Effort.jl}\footnote{\href{https://github.com/CosmologicalEmulators/Effort.jl}{https://github.com/CosmologicalEmulators/Effort.jl}} emulator is a novel computational tool designed for rapid and reliable analyses in the context of the EFTofLSS. Built on top of \texttt{pybird} and \texttt{velocileptors}, it shares computational backends with \texttt{Capse.jl}~\cite{bonici_capse}, providing compatibility with automatic differentiation engines and advanced inference frameworks like \texttt{Turing.jl}. \texttt{Effort.jl} integrates seamlessly with gradient-based samplers optimizers, such as NUTS and L-BFGS, to ensure efficient analyses even in high-dimensional spaces. Its design enables quick computation of galaxy clustering multipoles as functions of cosmological and nuisance parameters, making it particularly well-suited for full-shape analyses of galaxy power spectra.

Regarding its architecture, \texttt{Effort.jl} uses three multilayer perceptrons (MPLs), one for each of the multipole components, $P_{11}$, $P_{\rm loop}$, and $P_{\rm ct}$. For each MLP employed, we use 5 hidden layers with 64 coefficients. We use a training dataset with 20,000 samples, distributed according to the Latin Hypercube algorithm. We train our emulators using the ADAM optimizer, for around 1,000 epochs, starting from a learning rate of $10^{-4}$. After the training, \texttt{Effort.jl} is able to retrieve a multipole in around $40\,\mu\mathrm{s}$.

\begin{figure}
    \centering
\includegraphics[scale=0.8]{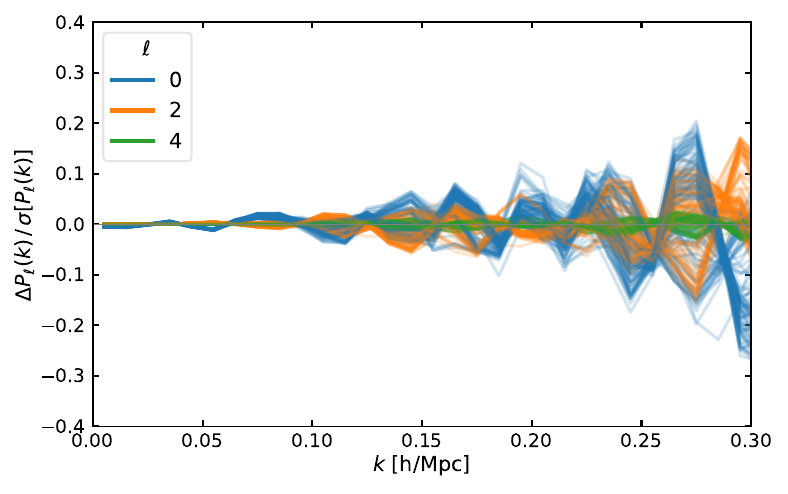}
    \caption{Emulator error on model predictions for the power spectrum multipoles as a fraction of the statistical uncertainty with an effective volume of $1000\,\mathrm{Gpc}^3$. The curves shown correspond to 100 points in parameter space sampled from a typical posterior for the case where $f_{\rm sat}=0.11$, $k_{\rm max}=0.2\,h\,\rm{Mpc}^{-1}$ and $\ell\in\{0,2,4\}$ (though results are  qualitatively the same for other values).}
    \label{fig:emulator_error}
\end{figure}

The accuracy of the emulator can be understood from Fig.~\ref{fig:emulator_error}; even in the quite extreme scenario of an effective volume of $1000$ $\mathrm{Gpc}^3$, the emulator error is just a fraction of the measurement error.

In order to employ \texttt{Effort.jl} in this analysis, we use the \texttt{JAX} version of the emulators\footnote{\href{https://github.com/CosmologicalEmulators/jaxeffort}{https://github.com/CosmologicalEmulators/jaxeffort}}. We do this for computational convenience and to more easily interface with \texttt{PocoMC}. We explicitly checked that \texttt{Effort.jl} and \texttt{jaxeffort} retrieve the same results, up to floating point precision.

\section{Analytic covariance for mock EFT data vector}
\label{appendix:analytic_gauss_covmat}
In section~\ref{sec:eft_mock_dvec} we put together a synthetic data vector by taking a perturbative EFT model for the redshift-space power spectrum and inducing non-linear damping by analytic means to resemble the effect of FoG. Though we were able to qualitatively reproduce the main features observed in the clustering of real galaxy samples, the power spectrum multipoles differ in detail from the actual clustering seen in data, and therefore also from that of simulations, which are designed to match the latter. We are therefore not able to use simulations to estimate the covariance matrix of our synthetic data vector. Instead, we use an analytic treatment, taking advatange of the fact that we know the anisotropic power spectrum exactly. 

In the absence of observational non-idealities, the disconnected part of the covariance of the power spectrum multipoles can be obtained as follows (see, e.g., Ref.~\cite{grieb_gaussian_2016}). First, we define the per-mode covariance,
\begin{equation}
    \sigma_{\ell_1 \ell_2}^2(k) = \frac{(2\ell_1 +1)(2\ell_2 +1)}{V_{\rm{s}}} \int_{-1}^1 \left[P(k,\mu) + P_{\rm{shot}}\right]^2 \mathcal{L}_{\ell_1} (\mu)\mathcal{L}_{\ell_2}(\mu) d\mu\,,
\end{equation}
where $V_{\rm{s}}$ is the volume of the survey; in our case case, this is a somewhat arbitrary choice, which we set to $V_{\rm{s}}=2000\, (2\,h^{-1}\,\rm{Gpc})^3$ in order to achieve fine precision while preserving the correct relative importance of different scales and multipoles.

But modes are typically grouped into bins in $k$ to increase the S/N and reduce the size of the covariance matrix. The bin-averaged power spectrum covariance can be obtained from the per-mode result as~\cite{grieb_gaussian_2016}
\begin{equation}
    \mathrm{Cov}\left[P^i_{\ell_1}, P^j_{\ell_2}\right] = \frac{2(2\pi)^4}{V^2_{k_i}} \delta_{ij} \int^{k_i + \Delta k/2}_{k_i - \Delta k/2} \sigma_{\ell_1 \ell_2}^2(k) k^2 dk\,,
\end{equation}
where the volume of the $i$-th $k$-space shell is $V_{k_i}=4\pi[(k_i+\Delta k/2)^3 - (k_i-\Delta k/2)^3 ]/3$, with $k_i$ labeling the bin center. The case at hand has the peculiarity that we are choosing and then fitting a synthetic data vector at the level of the power spectrum. This gives us the freedom to assert that the data vector is flat within our chosen bins, thus avoiding any need to rebin the data or the model prediction. We do still need to adjust the covariance, however, to account for the implicit averaging within $k$-bins. Under our assumption that the power spectrum is constant within bins, the covariance simplifies to
\begin{equation}
    \mathrm{Cov}\left[P^i_{\ell_1}, P^j_{\ell_2}\right] = (2\pi)^3 \delta_{ij} \frac{\sigma_{\ell_1 \ell_2}^2(k_i)}{V_{k_i}} \,.
\end{equation}

\section{Fitting the phenomenological FoG damping factor at fixed cosmology}\label{appendix:fitting_damping}
In this appendix, we explore whether the EFT model has sufficient flexibility to parametrize a phenomenological description of the FoG damping effect when the cosmology is fixed to the known truth and the nuisance parameters are given full freedom.  We will restrict our attention to the monopole and quadrupole only and the restricted EFT parametrization which varies only $\{b_1, b_2, b_s, \alpha_0, \alpha_2, \rm{SN}_0, \rm{SN}_2\}$ and sets all other nuisance parameters to zero --- this is the fiducial choice in the DESI 2024 full-shape analysis~\cite{collaborationDESI2024VII2024}. Adding additional parameters would no doubt improve the fits, but we choose to focus on this scenario that is more demanding for the model.

To test the flexibility of the EFT parametrization, we generate a redshift-space power spectrum with a rather arbitrary choice of nuisance parameters: $b_1=2, b_2=b_{s}=\alpha_0=\alpha_2=\rm{SN}_0=\rm{SN}_2=1$, cosmological parameters set to the base \texttt{Abacus}/Planck cosmology, and all others set to zero. Then we apply to this manifestly perturbative spectrum a Lorentzian damping factor characterized by a velocity dispersion of $\sigma_v=8\,h^{-1}\,\rm{Mpc}$. This rather high value for the satellite velocity dispersion makes it a conservative choice and ensures that $P_2(k_*)=0$ at $k_*\approx0.10h^{-1}\,\rm Mpc$, as is seen for satellite-satellite pairs in the simulation of \S\ref{sec:breakdown_of_Pk} above, as well as in other works in the literature~\cite{okumura_galaxy_2015}. 

Next, we define a Gaussian likelihood function for the EFT nuisance parameters given this mock data using the analytic disconnected covariances of appendix~\ref{appendix:analytic_gauss_covmat}, our synthetic data vector, and the full \texttt{velocileptors} model (since we are not varying the cosmology, the \texttt{velocileptors} tables only need to be computed once, making the model very cheap to compute). Finally, we maximize this likelihood while fixing cosmological parameters to the known truth for computational efficiency. We achieve this through simulated annealing (see \S\ref{sec:eft_mock_dvec} for a description of the method).

\begin{figure}
    \centering
\includegraphics[scale=0.74]{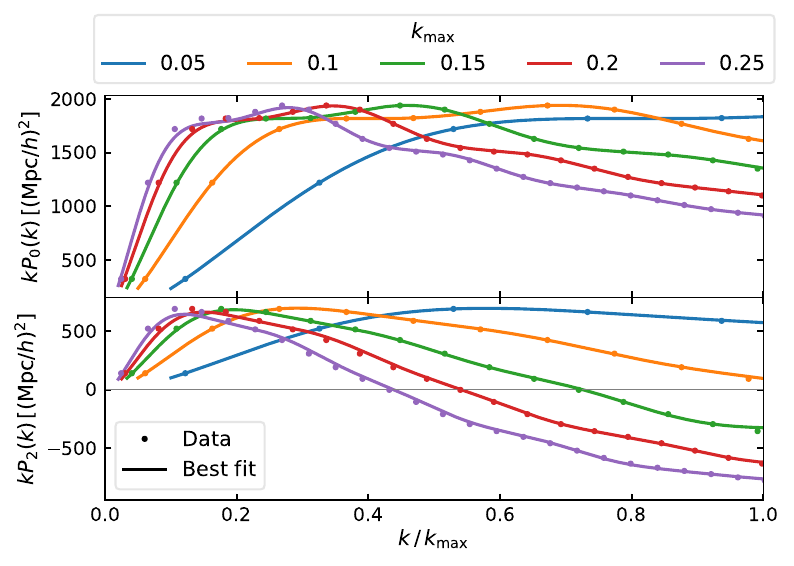}
    \caption{Ability of the EFT framework to fit synthetic data drawn from a composite model where FoG are included through a phenomenological damping factor rescaling a redshift-space power spectrum derived from the EFT parametrization. 
    We consider only the satellite-satellite contribution, assuming a velocity dispersion of $8\,h^{-1}\,\rm{Mpc}$ and obtaining a zero-crossing of the quadrupole at $k\approx 0.10\,h\,\rm{Mpc}^{-1}$. We compare the noiseless synthetic data (discrete points) against best fit-models (solid lines) obtained from different $k_{\rm max}$. The power spectrum monopole and quadrupole are shown in the upper and lower panels, respectively. Error bars are too small to be discernible, as they correspond to a volume of $1000\,(\mathrm{Gpc}/h)^3$.  This figure shows that when no external constraints are placed on the EFT parameters, the model has sufficient freedom to capture FoG damping even at $k$ significantly higher than the zero-crossing of the quadrupole --- that is, well into the non-perturbative regime. However, on such scales, the best-fit values of the parameters (see table~\ref{tab:bf_params_satonly_EFT}) are disfavored by typical priors and point to a breakdown of perturbativity. }
    \label{fig:sats_only_fits}
\end{figure}

Figure~\ref{fig:sats_only_fits} shows the monopole and quadrupole of the synthetic data vector as well as the best-fit models for various values of $k_{\rm max}$. Interestingly, when no constraints are imposed on the parameter values, the EFT model is flexible enough to provide a good fit to the data for $k_{\rm max}$ well beyond the zero-crossing scale. Note that this is true in spite of the fact that we are not using the full parameter freedom available at one-loop order. However, the best-fit parameters --- given in table~\ref{tab:bf_params_satonly_EFT} --- take on values that would be excluded by typical priors or be deemed unphysical on the grounds of perturbativity (see the main text for further details on this last point). 
\begin{table}[h]
    \centering
    \begin{tabular}{||c||c|c|c|c|c||}
        \hline
            $k_{\rm max}$ [$h^{-1}\,$Mpc] & 0.05 & 0.10 & 0.15 & 0.20 & 0.25\\
         \hline
         \hline
         $b_1$ & $1.00 \pm 0.02$ & $0.98 \pm 0.02$ & $0.93 \pm 0.03$ & $0.798 \pm 0.008$ & $0.735 \pm 0.006$  \\
         $b_2$ & $-4.8\pm 0.7$ & $-1.8\pm 0.7$ & $0.4\pm0.8$ & $-4.8\pm0.2$& $-4.28 \pm0.08$ \\
         $b_s$ & $1.5\pm 0.7$& $-2\pm1$ & $1.3\pm0.6$ & $6.1\pm0.1$ & $6.19\pm 0.04$ \\
         $\alpha_0$ & $-10 \pm 10$ & $11\pm 7$ & $9 \pm 2$& $3\pm1$ & $15\pm1$ \\
         $\alpha_2 \times10^{-2}$ & $-2 \pm 1$ & $-2.2\pm 0.1$& $-2.1 \pm 0.2$& $-0.9 \pm 0.03$ & $-0.96\pm 0.02$ \\
         $\rm{SN}_0 \times 10^{-3}$  & $2\pm1$ & $-1\pm 2$ &  $-4\pm 2$ &  $-2.3\pm 0.2$ & $-3.9\pm 0.1$ \\
         $\rm{SN}_2 \times 10^{-5}$ & $10\pm20$ & $3\pm1$ & $1.5\pm0.2$ & $0.58\pm0.01$& $0.364\pm0.002$ \\ \hline
         $\chi^2_{\nu}$ &$2.8\times10^{-4}$ & 0.15 & 2.4 & 12 & 38 \\
         \hline
    \end{tabular}
    \caption{Best-fit parameter values for the fits in figure~\ref{fig:sats_only_fits}, with uncertainties derived from the information matrix (for a volume this large, these are dominated by the model's identifiability issues; the contribution from cosmic variance can be neglected). Note that the input data vector has parameters $b_1=2, b_2=b_{s}=\alpha_0=\alpha_2=\rm{SN}_0=\rm{SN}_2=1$, and comprises a quadrupole that goes negative around $k\approx 0.10 \,h\,\rm{Mpc}^{-1}$. We see that when the nuisance parameters are given full freedom, the fit can be good beyond the zero-crossing scale, even in the limit considered here where the volume is $1000\,(h^{-1}\,\rm{Gpc})^3$. However, as we will see, this is for parameter values that are disfavored by typical priors or which are deemed to be unphysical --- note in particular the very high best-fit values for $\alpha_2$ which indicate a breakdown of perturbativity.\label{tab:bf_params_satonly_EFT}}
    \label{tab:}
\end{table}

\bibliographystyle{JHEP}
\bibliography{main} 

\providecommand{\href}[2]{#2}\begingroup\raggedright\begin{thebibliography}{10}

\bibitem{DESI}
{DESI Collaboration}, A.~{Aghamousa}, J.~{Aguilar}, et~al., {\it {The DESI Experiment Part I: Science,Targeting, and Survey Design}},  {\em ArXiv e-prints} (Oct., 2016) [\href{http://arxiv.org/abs/1611.00036}{{\tt arXiv:1611.00036}}].

\bibitem{laureijs2011euclid}
R.~Laureijs, J.~Amiaux, S.~Arduini, et~al., {\it Euclid definition study report},  2011.

\bibitem{Euclid:2024yrr}
{\bf Euclid} Collaboration, Y.~Mellier et~al., {\it {Euclid. I. Overview of the Euclid mission}},  \href{http://arxiv.org/abs/2405.13491}{{\tt arXiv:2405.13491}}.

\bibitem{kaiser_clustering_1987}
N.~Kaiser, {\it Clustering in real space and in redshift space},  {\em Monthly Notices of the Royal Astronomical Society} {\bf 227} (July, 1987) 1--21.

\bibitem{taruya_baryon_2010}
A.~Taruya, T.~Nishimichi, and S.~Saito, {\it Baryon {Acoustic} {Oscillations} in {2D}: {Modeling} {Redshift}-space {Power} {Spectrum} from {Perturbation} {Theory}},  June, 2010.
\newblock arXiv:1006.0699.

\bibitem{seljak_distribution_2011}
U.~Seljak and P.~McDonald, {\it Distribution function approach to redshift space distortions},  Sept., 2011.
\newblock arXiv:1109.1888.

\bibitem{reid_towards_2011}
B.~A. Reid and M.~White, {\it Towards an accurate model of the redshift space clustering of halos in the quasilinear regime},  May, 2011.
\newblock arXiv:1105.4165.

\bibitem{hand_extending_2017}
N.~Hand, U.~Seljak, F.~Beutler, and Z.~Vlah, {\it Extending the modeling of the anisotropic galaxy power spectrum to $k = 0.4 \, h\,\rm{Mpc}^{-1}$},  {\em J. Cosmol. Astropart. Phys.} {\bf 2017} (Oct., 2017) 009--009. arXiv:1706.02362 [astro-ph].

\bibitem{baumann_cosmological_2010}
D.~Baumann, A.~Nicolis, L.~Senatore, and M.~Zaldarriaga, {\it Cosmological {Non}-{Linearities} as an {Effective} {Fluid}},  Apr., 2010.
\newblock arXiv:1004.2488.

\bibitem{carrasco_effective_2012}
J.~J.~M. Carrasco, M.~P. Hertzberg, and L.~Senatore, {\it The {Effective} {Field} {Theory} of {Cosmological} {Large} {Scale} {Structures}},  Oct., 2012.
\newblock arXiv:1206.2926.

\bibitem{vlahLagrangianEffectiveField2015}
Z.~Vlah, M.~White, and A.~Aviles, {\it A {Lagrangian} effective field theory},  {\em Journal of Cosmology and Astroparticle Physics} {\bf 2015} (Sept., 2015) 014--014. arXiv:1506.05264 [astro-ph].

\bibitem{ivanov_effective_2022}
M.~M. Ivanov, {\it Effective {Field} {Theory} for {Large} {Scale} {Structure}},  Dec., 2022.
\newblock arXiv:2212.08488.

\bibitem{damico_cosmological_2020}
G.~D'Amico, J.~Gleyzes, N.~Kokron, et~al., {\it The {Cosmological} {Analysis} of the {SDSS}/{BOSS} data from the {Effective} {Field} {Theory} of {Large}-{Scale} {Structure}},  {\em Journal of Cosmology and Astroparticle Physics} {\bf 2020} (May, 2020) 005--005. arXiv:1909.05271 [astro-ph, physics:gr-qc, physics:hep-ph, physics:hep-th].

\bibitem{ivanov_cosmological_2020}
M.~M. Ivanov, M.~Simonović, and M.~Zaldarriaga, {\it Cosmological {Parameters} from the {BOSS} {Galaxy} {Power} {Spectrum}},  {\em Journal of Cosmology and Astroparticle Physics} {\bf 2020} (May, 2020) 042--042. arXiv:1909.05277 [astro-ph, physics:gr-qc, physics:hep-ph].

\bibitem{ivanov_cosmological_2021}
M.~M. Ivanov, {\it Cosmological constraints from the power spectrum of {eBOSS} emission line galaxies},  {\em Physical Review D} {\bf 104} (Nov., 2021) 103514. arXiv:2106.12580 [astro-ph].

\bibitem{philcox_boss_2022}
O.~H.~E. Philcox and M.~M. Ivanov, {\it The {BOSS} {DR12} {Full}-{Shape} {Cosmology}: \${\textbackslash}{Lambda}\${CDM} {Constraints} from the {Large}-{Scale} {Galaxy} {Power} {Spectrum} and {Bispectrum} {Monopole}},  {\em Physical Review D} {\bf 105} (Feb., 2022) 043517. arXiv:2112.04515 [astro-ph, physics:hep-ex].

\bibitem{chen_new_2022}
S.-F. Chen, Z.~Vlah, and M.~White, {\it A new analysis of galaxy 2-point functions in the {BOSS} survey, including full-shape information and post-reconstruction {BAO}},  {\em J. Cosmol. Astropart. Phys.} {\bf 2022} (Feb., 2022) 008.

\bibitem{ivanov_full-shape_2024}
M.~M. Ivanov, A.~Obuljen, C.~Cuesta-Lazaro, and M.~W. Toomey, {\it Full-shape analysis with simulation-based priors: cosmological parameters and the structure growth anomaly},  Sept., 2024.
\newblock arXiv:2409.10609.

\bibitem{collaborationDESI2024VII2024}
D.~Collaboration, A.~G. Adame, J.~Aguilar, et~al., {\it {DESI} 2024 {VII}: {Cosmological} {Constraints} from the {Full}-{Shape} {Modeling} of {Clustering} {Measurements}},  Nov., 2024.
\newblock arXiv:2411.12022.

\bibitem{dodelson_12_2021}
S.~Dodelson and F.~Schmidt, {\it 12 - {Growth} of structure: beyond linear theory},  in {\em Modern {Cosmology} ({Second} {Edition})} (S.~Dodelson and F.~Schmidt, eds.), pp.~325--372.
\newblock Academic Press, Jan., 2021.

\bibitem{nishimichi_blinded_2020}
T.~Nishimichi, G.~D'Amico, M.~M. Ivanov, et~al., {\it Blinded challenge for precision cosmology with large-scale structure: results from effective field theory for the redshift-space galaxy power spectrum},  {\em Phys. Rev. D} {\bf 102} (Dec., 2020) 123541. arXiv:2003.08277 [astro-ph].

\bibitem{maus_analysis_2024}
M.~Maus, S.~Chen, M.~White, et~al., {\it An analysis of parameter compression and full-modeling techniques with {Velocileptors} for {DESI} 2024 and beyond},  Apr., 2024.
\newblock arXiv:2404.07312 [astro-ph].

\bibitem{taule_two-loop_2023}
P.~Taule and M.~Garny, {\it The two-loop power spectrum in redshift space},  {\em Journal of Cosmology and Astroparticle Physics} {\bf 2023} (Nov., 2023) 078. arXiv:2308.07379 [astro-ph, physics:hep-ph].

\bibitem{tegmark_cosmological_2006}
M.~Tegmark, D.~J. Eisenstein, M.~A. Strauss, et~al., {\it Cosmological constraints from the {SDSS} luminous red galaxies},  {\em Physical Review D} {\bf 74} (Dec., 2006) 123507. Publisher: American Physical Society.

\bibitem{reid_luminous_2009}
B.~A. Reid, D.~N. Spergel, and P.~Bode, {\it Luminous {Red} {Galaxy} {Halo} {Density} {Field} {Reconstruction} and {Application} to {Large} {Scale} {Structure} {Measurements}},  {\em The Astrophysical Journal} {\bf 702} (Sept., 2009) 249--265. arXiv:0811.1025 [astro-ph].

\bibitem{ivanov_cosmological_2022}
M.~M. Ivanov, O.~H.~E. Philcox, M.~Simonović, et~al., {\it Cosmological constraints without fingers of {God}},  {\em Physical Review D} {\bf 105} (Feb., 2022) 043531. arXiv:2110.00006 [astro-ph].

\bibitem{damico_taming_2021}
G.~D'Amico, L.~Senatore, P.~Zhang, and T.~Nishimichi, {\it Taming redshift-space distortion effects in the {EFTofLSS} and its application to data},  Sept., 2021.

\bibitem{schmittfull_modeling_2021}
M.~Schmittfull, M.~Simonović, M.~M. Ivanov, et~al., {\it Modeling galaxies in redshift space at the field level},  {\em Journal of Cosmology and Astroparticle Physics} {\bf 2021} (May, 2021) 059. Publisher: IOP ADS Bibcode: 2021JCAP...05..059S.

\bibitem{coilDEEP2GalaxyRedshift2008}
A.~L. Coil, J.~A. Newman, D.~Croton, et~al., {\it The {DEEP2} {Galaxy} {Redshift} {Survey}: {Color} and {Luminosity} {Dependence} of {Galaxy} {Clustering} at z{\textasciitilde}1},  {\em The Astrophysical Journal} {\bf 672} (Jan., 2008) 153--176. arXiv:0708.0004 [astro-ph].

\bibitem{mohammadVIMOSPublicExtragalactic2018}
F.~G. Mohammad, B.~R. Granett, L.~Guzzo, et~al., {\it The {VIMOS} {Public} {Extragalactic} {Redshift} {Survey} ({VIPERS}): {An} unbiased estimate of the growth rate of structure at <z>=0.85 using the clustering of luminous blue galaxies},  {\em Astronomy \& Astrophysics} {\bf 610} (Feb., 2018) A59. arXiv:1708.00026 [astro-ph].

\bibitem{hangGalaxyMassAssembly2022}
Q.~Hang, J.~A. Peacock, S.~Alam, et~al., {\it Galaxy and {Mass} {Assembly} ({GAMA}): {Probing} galaxy-group correlations in redshift space with the halo streaming model},  {\em Monthly Notices of the Royal Astronomical Society} {\bf 517} (Oct., 2022) 374--392. arXiv:2206.05065 [astro-ph].

\bibitem{mergulhaoEffectiveFieldTheory2023}
T.~Mergulhão, H.~Rubira, and R.~Voivodic, {\it The {Effective} {Field} {Theory} of {Large}-{Scale} {Structure} and {Multi}-tracer {II}: redshift space and realistic tracers},  June, 2023.
\newblock arXiv:2306.05474 [astro-ph].

\bibitem{sunyaev_observations_1972}
R.~A. Sunyaev and Y.~B. Zeldovich, {\it The {Observations} of {Relic} {Radiation} as a {Test} of the {Nature} of {X}-{Ray} {Radiation} from the {Clusters} of {Galaxies}},  {\em Comments on Astrophysics and Space Physics} {\bf 4} (Nov., 1972) 173.

\bibitem{ade_simons_2019}
P.~Ade, J.~Aguirre, Z.~Ahmed, et~al., {\it The {Simons} {Observatory}: science goals and forecasts},  {\em Journal of Cosmology and Astroparticle Physics} {\bf 2019} (Feb., 2019) 056.

\bibitem{abazajian_cmb-s4_2016}
K.~N. Abazajian, P.~Adshead, Z.~Ahmed, et~al., {\it {CMB}-{S4} {Science} {Book}, {First} {Edition}},  Oct., 2016.
\newblock arXiv:1610.02743 [astro-ph, physics:gr-qc, physics:hep-ph, physics:hep-th].

\bibitem{sanchezClusteringGalaxiesCompleted2017}
A.~G. Sánchez, R.~Scoccimarro, M.~Crocce, et~al., {\it The clustering of galaxies in the completed {SDSS}-{III} {Baryon} {Oscillation} {Spectroscopic} {Survey}: {Cosmological} implications of the configuration-space clustering wedges},  {\em Monthly Notices of the Royal Astronomical Society} {\bf 464} (Jan., 2017) 1640--1658.

\bibitem{peacock_power_1992}
J.~A. Peacock and M.~J. West, {\it The power spectrum of {Abell} cluster correlations},  {\em Monthly Notices of the Royal Astronomical Society} {\bf 259} (Dec., 1992) 494--504.

\bibitem{percival_testing_2009}
W.~J. Percival and M.~White, {\it Testing cosmological structure formation using redshift-space distortions},  {\em Monthly Notices of the Royal Astronomical Society} {\bf 393} (Feb., 2009) 297--308. arXiv:0808.0003 [astro-ph].

\bibitem{chen_consistent_2020}
S.-F. Chen, Z.~Vlah, and M.~White, {\it Consistent {Modeling} of {Velocity} {Statistics} and {Redshift}-{Space} {Distortions} in {One}-{Loop} {Perturbation} {Theory}},  {\em J. Cosmol. Astropart. Phys.} {\bf 2020} (July, 2020) 062--062. arXiv:2005.00523 [astro-ph].

\bibitem{okumura_galaxy_2015}
T.~Okumura, N.~Hand, U.~Seljak, et~al., {\it Galaxy power spectrum in redshift space: combining perturbation theory with the halo model},  {\em Phys. Rev. D} {\bf 92} (Nov., 2015) 103516. arXiv:1506.05814 [astro-ph].

\bibitem{peebles_textbook}
P.~J.~E. {Peebles}, {\em {The large-scale structure of the universe}}.
\newblock 1980.

\bibitem{planck_18_params}
P.~Collaboration, N.~Aghanim, Y.~Akrami, et~al., {\it Planck 2018 results. {VI}. {Cosmological} parameters},  July, 2018.

\bibitem{maksimova_span_2021}
N.~A. Maksimova, L.~H. Garrison, D.~J. Eisenstein, et~al., {\it Abacussummit: a massive set of high-accuracy, high-resolution \textit{{N}}-body simulations},  {\em Monthly Notices of the Royal Astronomical Society} {\bf 508} (Oct., 2021) 4017--4037.

\bibitem{dawson_baryon_2013}
K.~S. Dawson, D.~J. Schlegel, C.~P. Ahn, et~al., {\it The {Baryon} {Oscillation} {Spectroscopic} {Survey} of {SDSS}-{III}},  {\em The Astronomical Journal} {\bf 145} (Jan., 2013) 10. arXiv:1208.0022 [astro-ph].

\bibitem{collaboration_desi_2024}
D.~Collaboration, A.~G. Adame, J.~Aguilar, et~al., {\it {DESI} 2024 {VI}: {Cosmological} {Constraints} from the {Measurements} of {Baryon} {Acoustic} {Oscillations}},  Apr., 2024.
\newblock arXiv:2404.03002.

\bibitem{yuan_span_2022}
S.~Yuan, L.~H. Garrison, B.~Hadzhiyska, et~al., {\it Abacushod: a highly efficient extended multitracer {HOD} framework and its application to {BOSS} and {eBOSS} data},  {\em Monthly Notices of the Royal Astronomical Society} {\bf 510} (Jan., 2022) 3301--3320.

\bibitem{zheng_galaxy_2007}
Z.~Zheng, A.~L. Coil, and I.~Zehavi, {\it Galaxy {Evolution} from {Halo} {Occupation} {Distribution} {Modeling} of {DEEP2} and {SDSS} {Galaxy} {Clustering}},  {\em The Astrophysical Journal} {\bf 667} (Oct., 2007) 760. Publisher: IOP Publishing.

\bibitem{guoVelocityBiasSmall2015}
H.~Guo, Z.~Zheng, I.~Zehavi, et~al., {\it Velocity {Bias} from the {Small} {Scale} {Clustering} of {SDSS}-{III} {BOSS} {Galaxies}},  {\em Monthly Notices of the Royal Astronomical Society} {\bf 446} (Jan., 2015) 578--594. arXiv:1407.4811 [astro-ph].

\bibitem{leauthaudStripe82Massive2016}
A.~Leauthaud, K.~Bundy, S.~Saito, et~al., {\it The {Stripe} 82 {Massive} {Galaxy} {Project} – {II}. {Stellar} mass completeness of spectroscopic galaxy samples from the {Baryon} {Oscillation} {Spectroscopic} {Survey}},  {\em Monthly Notices of the Royal Astronomical Society} {\bf 457} (Apr., 2016) 4021--4037.

\bibitem{guoIncompleteConditionalStellar2018}
H.~Guo, X.~Yang, and Y.~Lu, {\it The {Incomplete} {Conditional} {Stellar} {Mass} {Function}: {Unveiling} the {Stellar} {Mass} {Functions} of {Galaxies} at 0.1 {\textless} {Z} {\textless} 0.8 from {BOSS} {Observations}},  {\em The Astrophysical Journal} {\bf 858} (May, 2018) 30. Publisher: The American Astronomical Society.

\bibitem{avila_completed_2020}
S.~Avila, V.~Gonzalez-Perez, F.~G. Mohammad, et~al., {\it The {Completed} {SDSS}-{IV} extended {Baryon} {Oscillation} {Spectroscopic} {Survey}: exploring the {Halo} {Occupation} {Distribution} model for {Emission} {Line} {Galaxies}},  {\em Monthly Notices of the Royal Astronomical Society} {\bf 499} (Nov., 2020) 5486--5507. arXiv:2007.09012 [astro-ph].

\bibitem{chen_redshift-space_2021}
S.-F. Chen, Z.~Vlah, E.~Castorina, and M.~White, {\it Redshift-{Space} {Distortions} in {Lagrangian} {Perturbation} {Theory}},  {\em J. Cosmol. Astropart. Phys.} {\bf 2021} (Mar., 2021) 100. arXiv:2012.04636 [astro-ph].

\bibitem{maus_comparison_2024}
M.~Maus, Y.~Lai, H.~E. Noriega, et~al., {\it A comparison of effective field theory models of redshift space galaxy power spectra for {DESI} 2024 and future surveys},  Apr., 2024.
\newblock arXiv:2404.07272 [astro-ph].

\bibitem{kaiser_spatial_1984}
N.~Kaiser, {\it On the spatial correlations of {Abell} clusters.},  {\em The Astrophysical Journal} {\bf 284} (Sept., 1984) L9--L12. ADS Bibcode: 1984ApJ...284L...9K.

\bibitem{abidi_cubic_2018}
M.~M. Abidi and T.~Baldauf, {\it Cubic {Halo} {Bias} in {Eulerian} and {Lagrangian} {Space}},  {\em Journal of Cosmology and Astroparticle Physics} {\bf 2018} (July, 2018) 029--029. arXiv:1802.07622 [astro-ph].

\bibitem{Lewis:1999bs}
A.~Lewis, A.~Challinor, and A.~Lasenby, {\it {Efficient computation of CMB anisotropies in closed FRW models}},  {\em \apj} {\bf 538} (2000) 473--476, [\href{http://xxx.lanl.gov/abs/astro-ph/9911177}{{\tt astro-ph/9911177}}].

\bibitem{lesgourgues_cosmic_2011}
J.~Lesgourgues, {\it The {Cosmic} {Linear} {Anisotropy} {Solving} {System} ({CLASS}) {I}: {Overview}},  May, 2011.
\newblock arXiv:1104.2932 [astro-ph].

\bibitem{damicoLimitsPrimordialNonGaussianities2022}
G.~D'Amico, M.~Lewandowski, L.~Senatore, and P.~Zhang, {\it Limits on primordial non-{Gaussianities} from {BOSS} galaxy-clustering data},  Jan., 2022.
\newblock Publication Title: arXiv e-prints ADS Bibcode: 2022arXiv220111518D.

\bibitem{bragancaPeekingNextDecade2023}
D.~Bragança, Y.~Donath, L.~Senatore, and H.~Zheng, {\it Peeking into the next decade in {Large}-{Scale} {Structure} {Cosmology} with its {Effective} {Field} {Theory}},  July, 2023.
\newblock Publication Title: arXiv e-prints ADS Bibcode: 2023arXiv230704992B.

\bibitem{zhangHODinformedPriorEFTbased2024}
H.~Zhang, M.~Bonici, G.~D'Amico, et~al., {\it {HOD}-informed prior for {EFT}-based full-shape analyses of {LSS}},  Sept., 2024.
\newblock arXiv:2409.12937.

\bibitem{boniciEffortFastDifferentiable2025}
M.~Bonici, G.~D'Amico, J.~Bel, and C.~Carbone, {\it Effort: a fast and differentiable emulator for the {Effective} {Field} {Theory} of the {Large} {Scale} {Structure} of the {Universe}},  Jan., 2025.
\newblock arXiv:2501.04639 [astro-ph].

\bibitem{karamanis2022pocomc}
M.~Karamanis, D.~Nabergoj, F.~Beutler, et~al., {\it poco{MC}: {A} {P}ython package for accelerated {B}ayesian inference in astronomy and cosmology},  {\em arXiv preprint arXiv:2207.05660} (2022).

\bibitem{karamanis_accelerating_2022}
M.~Karamanis, F.~Beutler, J.~A. Peacock, et~al., {\it Accelerating astronomical and cosmological inference with {Preconditioned} {Monte} {Carlo}},  {\em Monthly Notices of the Royal Astronomical Society} {\bf 516} (Sept., 2022) 1644--1653. arXiv:2207.05652 [astro-ph, physics:physics].

\bibitem{taule_two-loop_nodate}
P.~Taule and M.~Garny, {\it The two-loop power spectrum in redshift space}, .

\bibitem{mergulhao_effective_2022}
T.~Mergulhão, H.~Rubira, R.~Voivodic, and L.~R. Abramo, {\it The {Effective} {Field} {Theory} of {Large}-{Scale} {Structure} and {Multi}-tracer},  {\em Journal of Cosmology and Astroparticle Physics} {\bf 2022} (Apr., 2022) 021. arXiv:2108.11363 [astro-ph, physics:gr-qc, physics:hep-ph].

\bibitem{ebina_cosmology_2024}
H.~Ebina and M.~White, {\it Cosmology before noon with multiple galaxy populations},  Jan., 2024.
\newblock arXiv:2401.13166 [astro-ph].

\bibitem{hadzhiyskaCosmology6Parameters2023}
B.~Hadzhiyska, K.~Wolz, S.~Azzoni, et~al., {\it Cosmology with 6 parameters in the {Stage}-{IV} era: efficient marginalisation over nuisance parameters},  {\em The Open Journal of Astrophysics} {\bf 6} (July, 2023) 10.21105/astro.2301.11895. arXiv:2301.11895 [astro-ph].

\bibitem{dresslerGalaxyMorphologyRich1980}
A.~Dressler, {\it Galaxy morphology in rich clusters - {Implications} for the formation and evolution of galaxies},  {\em The Astrophysical Journal} {\bf 236} (Mar., 1980) 351.

\bibitem{madgwick2dFGalaxyRedshift2003}
D.~S. Madgwick, E.~Hawkins, O.~Lahav, et~al., {\it The {2dF} {Galaxy} {Redshift} {Survey}: galaxy clustering per spectral type},  {\em Monthly Notices of the Royal Astronomical Society} {\bf 344} (Sept., 2003) 847--856. arXiv:astro-ph/0303668.

\bibitem{singh_fundamental_2021}
S.~Singh, B.~Yu, and U.~Seljak, {\it Fundamental {Plane} of {BOSS} galaxies: {Correlations} with galaxy properties, density field and impact on {RSD} measurements},  {\em Monthly Notices of the Royal Astronomical Society} {\bf 501} (Jan., 2021) 4167--4183. arXiv:2001.07700 [astro-ph].

\bibitem{lemboCMBLensingReconstruction2022}
M.~Lembo, G.~Fabbian, J.~Carron, and A.~Lewis, {\it {CMB} lensing reconstruction biases from masking extragalactic sources},  {\em Physical Review D} {\bf 106} (July, 2022) 023525. arXiv:2109.13911 [astro-ph].

\bibitem{surraoAccurateEstimationAngular2023}
K.~M. Surrao, O.~H.~E. Philcox, and J.~C. Hill, {\it Accurate estimation of angular power spectra for maps with correlated masks},  {\em Physical Review D} {\bf 107} (Apr., 2023) 083521. arXiv:2302.05436 [astro-ph].

\bibitem{the_simons_observatory_collaboration_simons_2019}
T.~S.~O. Collaboration, P.~Ade, J.~Aguirre, et~al., {\it The {Simons} {Observatory}: {Science} goals and forecasts},  {\em Journal of Cosmology and Astroparticle Physics} {\bf 2019} (Feb., 2019) 056--056. arXiv:1808.07445 [astro-ph].

\bibitem{zubeldia_cosmocnc_2024}
I.~Zubeldia and B.~Bolliet, {\it cosmocnc: {A} fast, flexible, and accurate galaxy cluster number count likelihood code for cosmology},  Mar., 2024.
\newblock arXiv:2403.09589 [astro-ph].

\bibitem{merloniEROSITAScienceBook2012}
A.~Merloni, P.~Predehl, W.~Becker, et~al., {\it {eROSITA} {Science} {Book}: {Mapping} the {Structure} of the {Energetic} {Universe}},  Sept., 2012.
\newblock Publication Title: arXiv e-prints ADS Bibcode: 2012arXiv1209.3114M.

\bibitem{eggemeierBoostingGalaxyClustering2025}
A.~Eggemeier, N.~Lee, R.~Scoccimarro, et~al., {\it Boosting galaxy clustering analyses with non-perturbative modelling of redshift-space distortions},  Jan., 2025.
\newblock arXiv:2501.18597 [astro-ph] version: 1.

\bibitem{maus_comparison_2023}
M.~Maus, S.-F. Chen, and M.~White, {\it A comparison of template vs. direct model fitting for redshift-space distortions in {BOSS}},  {\em J. Cosmol. Astropart. Phys.} {\bf 2023} (June, 2023) 005. Publisher: IOP Publishing.

\bibitem{simon_consistency_2023}
T.~Simon, P.~Zhang, V.~Poulin, and T.~L. Smith, {\it Consistency of effective field theory analyses of the {BOSS} power spectrum},  {\em Phys. Rev. D} {\bf 107} (June, 2023) 123530. arXiv:2208.05929 [astro-ph, physics:hep-ph, physics:hep-th].

\bibitem{holm_bayesian_2023}
E.~B. Holm, L.~Herold, T.~Simon, et~al., {\it Bayesian and frequentist investigation of prior effects in {EFTofLSS} analyses of full-shape {BOSS} and {eBOSS} data},  Sept., 2023.
\newblock arXiv:2309.04468 [astro-ph, physics:hep-ph].

\bibitem{ivanovFullshapeAnalysisSimulationbased2024}
M.~M. Ivanov, C.~Cuesta-Lazaro, S.~Mishra-Sharma, et~al., {\it Full-shape analysis with simulation-based priors: constraints on single field inflation from {BOSS}},  Feb., 2024.
\newblock arXiv:2402.13310 [astro-ph, physics:hep-ph, physics:hep-th].

\bibitem{eisenstein_improving_2007}
D.~J. Eisenstein, H.-j. Seo, E.~Sirko, and D.~Spergel, {\it Improving {Cosmological} {Distance} {Measurements} by {Reconstruction} of the {Baryon} {Acoustic} {Peak}},  {\em The Astrophysical Journal} {\bf 664} (Aug., 2007) 675--679. arXiv:astro-ph/0604362.

\bibitem{guachalla_velocity_2024}
B.~Guachalla~Ried, E.~Schaan, B.~Hadzhiyska, and S.~Ferraro, {\it Velocity reconstruction in the era of {DESI} and {Rubin} (part {I}): {Exploring} spectroscopic, photometric \& hybrid samples},  {\em Physical Review D} {\bf 109} (May, 2024) 103533. arXiv:2312.12435 [astro-ph].

\bibitem{yu_rsd_2023}
B.~Yu, U.~Seljak, Y.~Li, and S.~Singh, {\it {RSD} measurements from {BOSS} galaxy power spectrum using the halo perturbation theory model},  {\em J. Cosmol. Astropart. Phys.} {\bf 2023} (Apr., 2023) 057. arXiv:2211.16794 [astro-ph].

\bibitem{hand_optimal_2017}
N.~Hand, Y.~Li, Z.~Slepian, and U.~Seljak, {\it An optimal {FFT}-based anisotropic power spectrum estimator},  Apr., 2017.
\newblock arXiv:1704.02357 [astro-ph].

\bibitem{sefusatti_accurate_2016}
E.~Sefusatti, M.~Crocce, R.~Scoccimarro, and H.~Couchman, {\it Accurate {Estimators} of {Correlation} {Functions} in {Fourier} {Space}},  {\em Mon. Not. R. Astron. Soc.} {\bf 460} (Aug., 2016) 3624--3636. arXiv:1512.07295 [astro-ph].

\bibitem{beutler_clustering_2017}
F.~Beutler, H.-J. Seo, S.~Saito, et~al., {\it The clustering of galaxies in the completed {SDSS}-{III} {Baryon} {Oscillation} {Spectroscopic} {Survey}: {Anisotropic} galaxy clustering in {Fourier}-space},  {\em Mon. Not. R. Astron. Soc.} {\bf 466} (Apr., 2017) 2242--2260. arXiv:1607.03150 [astro-ph].

\bibitem{alcock_evolution_1979}
C.~Alcock and B.~Paczyński, {\it An evolution free test for non-zero cosmological constant},  {\em Nature} {\bf 281} (Oct., 1979) 358--359. Publisher: Nature Publishing Group.

\bibitem{chudaykin_non-linear_2020}
A.~Chudaykin, M.~M. Ivanov, O.~H.~E. Philcox, and M.~Simonović, {\it Non-linear perturbation theory extension of the {Boltzmann} code {CLASS}},  {\em Phys. Rev. D} {\bf 102} (Sept., 2020) 063533. arXiv:2004.10607 [astro-ph].

\bibitem{bonici_capse}
M.~{Bonici}, F.~{Bianchini}, and J.~{Ruiz-Zapatero}, {\it {Capse.jl: efficient and auto-differentiable CMB power spectra emulation}},  {\em The Open Journal of Astrophysics} {\bf 7} (Jan., 2024) 10, [\href{http://arxiv.org/abs/2307.14339}{{\tt arXiv:2307.14339}}].

\bibitem{grieb_gaussian_2016}
J.~N. Grieb, A.~G. Sánchez, S.~Salazar-Albornoz, and C.~Dalla Vecchia, {\it Gaussian covariance matrices for anisotropic galaxy clustering measurements},  {\em Monthly Notices of the Royal Astronomical Society} {\bf 457} (Apr., 2016) 1577--1592.

\end{thebibliography}\endgroup

\end{document}